\documentclass[a4paper,12pt]{article}
\usepackage{amsmath}
\usepackage{amssymb}
\usepackage{fullpage}
\usepackage{verbatim}

\usepackage[T1,OT1]{fontenc}

\usepackage{graphicx}

\allowdisplaybreaks[3]

\usepackage[hang]{footmisc}
\usepackage[compress,square,numbers]{natbib}
\usepackage[colorlinks,linktocpage,linkcolor=blue,citecolor=blue,urlcolor=blue]{hyperref}

\renewcommand{\d}{\mathrm{d}}
\newcommand{\e}{\mathrm{e}}
\newcommand{\w}{\wedge}

\newcommand{\nl}{\notag \\ &\quad\,}
\newcommand{\nll}{\notag \\ &}

\begin{document}

\numberwithin{equation}{section}

\thispagestyle{empty}

\vspace*{1cm}

\begin{center}

{\LARGE \bf O-plane Backreaction and Scale Separation}

\vspace{0.5cm}

{\LARGE \bf in Type IIA Flux Vacua}

\vspace{1.5cm}
{\large Daniel Junghans}\\

\vspace{1cm}
Institut f{\"{u}}r Theoretische Physik, Ruprecht-Karls-Universit{\"{a}}t Heidelberg,\\
Philosophenweg 19, 69120 Heidelberg, Germany\\

\vspace{1cm}
{\upshape\ttfamily junghans@thphys.uni-heidelberg.de}\\

\vspace{1.5cm}

\begin{abstract}
\noindent 

We construct AdS$_4$ flux vacua of type IIA string theory in the supergravity (large volume, small $g_s$) regime, including the backreaction of O6-planes. Our solutions are the localized versions of the smeared solutions on Calabi-Yau orientifolds studied by DeWolfe, Giryavets, Kachru and Taylor and in other works.
We find that the O-plane backreaction in these solutions generates warping, a varying dilaton, non-closed RR field strengths and internal curvature.
Just like their smeared counterparts, the localized solutions admit stabilized moduli, a parametric control over string corrections and a parametric separation between the AdS and KK scales. Our explicit expressions furthermore make precise the common lore that smeared solutions should approximate the exact ones in the large-volume, small-$g_s$ limit.
Finally, our solutions appear to violate a recent swampland conjecture about an absence of scale separation in supersymmetric AdS vacua. We make a simple observation explaining why this happens in these solutions in contrast to most other AdS solutions in string theory.

\end{abstract}

\end{center}

\newpage

\tableofcontents

\section{Introduction}

Type IIA string theory compactified on Calabi-Yau orientifolds is known to have AdS flux vacua with a number of very attractive properties: First, all moduli are stabilized at tree-level (i.e., in the classical, two-derivative supergravity approximation) using only fluxes. Second, potentially dangerous string corrections to the supergravity equations are parametrically suppressed and thus well-controlled. Third, there is a parametric separation between the AdS curvature scale and the KK scale, making the vacua truly 4-dimensional.
Supersymmetric solutions with these properties were found by DeWolfe, Giryavets, Kachru and Taylor (DGKT) in \cite{DeWolfe:2005uu} (see also \cite{Derendinger:2004jn, Camara:2005dc, Acharya:2006ne}). Furthermore, there are non-supersymmetric solutions with the same properties \cite{Camara:2005dc, Narayan:2010em, Marchesano:2019hfb}.\footnote{Note that there are similar AdS vacua with tree-level moduli stabilization and scale separation in type IIB, but they require compactifying on non-Calabi-Yau orientifolds \cite{Caviezel:2008ik, Caviezel:2009tu, Petrini:2013ika}.}

A recurring criticism in the literature is that the DGKT solutions are not trustworthy because they were obtained in an approximation where the O6-planes are smeared, i.e., their tension and charge densities are assumed to be uniformly distributed over the compact space for simplicity. The DGKT solutions do therefore not solve the 10d equations of motion, where the O6-planes appear as localized codimension-3 objects (see, e.g., \cite{Douglas:2010rt, Saracco:2012wc, McOrist:2012yc, Lust:2019zwm} for comments along these lines).\footnote{See also \cite{Blaback:2010sj} for a discussion of smearing in Minkowski flux compactifications.} In the true solutions, the O-plane backreaction is expected to generate gradients in the various supergravity fields, which are not captured in the smeared approximation.
Over the past years, several works studied aspects of these backreaction effects in the DGKT setup \cite{Acharya:2006ne, Saracco:2012wc, Gautason:2015tig}. However, exact solutions including the full O6-plane backreaction have so far not been found. Consequently, it has remained unclear whether such solutions, if existent at all, share the intriguing features of the smeared ones.

The purpose of this paper is to close this gap. In particular, we will explicitly construct the exact solutions corresponding to the smeared DGKT solutions, including the backreaction of the O6-planes. Our analysis equally applies to the non-supersymmetric AdS solutions of \cite{Camara:2005dc, Narayan:2010em, Marchesano:2019hfb}, which arise in the same setting of type IIA Calabi-Yau orientifolds. We will work in the usual regime where string theory is described by 10d supergravity. This is the case at large internal volumes and small $g_s$, which in the DGKT setup corresponds to the regime $n\gg 1$. Here, $n$ is a parameter which is proportional to the $F_4$ flux numbers. 

As we will see below, the exact solution for the 10d fields at large $n$ is given by the corresponding smeared solution plus $1/\sqrt{n}$ corrections, which capture the backreaction effects and are determined by simple Poisson equations.
Our solution breaks down close to the O6-planes, where the backreaction effects and string corrections become locally relevant.
However, we will argue that, at large $n$, such local corrections are parametrically suppressed in the 4d scalar potential and therefore irrelevant for the low-energy physics. The exact scalar potential therefore agrees with the corresponding smeared expression, up to subleading terms.
Our results thus confirm and make precise the common lore that backreaction becomes negligible at large volumes and small $g_s$.
Note that this implies that the backreacted solutions have the same attractive features as the smeared ones. In particular,
the backreaction does not seem to destabilize the moduli or change the ratio between the AdS and KK scales.

To the best of our knowledge, our solutions are the first ones in the literature on type II flux compactifications where the backreaction of intersecting sources is taken into account.\footnote{F-theory models with 7-branes are important exceptions.} Our method is rather general and should therefore be applicable beyond the particular case of IIA Calabi-Yau orientifolds studied in this paper. As we will explain further below, the key insight is that the backreaction of branes or O-planes is determined by simple linear equations in the limit of large volumes and small $g_s$.
Our method is therefore expected to work for many flux compactifications admitting such a limit.

Our results are also interesting from the point of view of the recent swampland program. In particular, it was recently conjectured that string theory does not admit supersymmetric AdS vacua with AdS/KK scale separation in limits where the vacuum energy goes to zero (strong AdS distance conjecture) \cite{Lust:2019zwm}. The DGKT vacua with $n\gg 1$ appear to be counter-examples to this conjecture. This was already noted in \cite{Lust:2019zwm} but
suspected there to be an artifact of the smearing. However, our results show that the DGKT vacua correspond to genuine solutions of the type IIA supergravity equations and should therefore be taken seriously as potential examples of scale separation in string theory.
This suggests that whether or not an AdS vacuum exhibits scale separation is not necessarily determined by its supersymmetry.

Nevertheless, it is true that many (supersymmetric and non-supersymmetric) AdS solutions in string theory do not have any scale separation. One may therefore wonder what makes the DGKT vacua and their non-supersymmetric cousins so special.
In this paper, we will not attempt to determine the necessary and sufficient conditions for AdS/KK scale separation in full generality.
However, we will argue that a simple observation plausibly explains this behavior. Indeed, we will show that, due to the scale invariance of classical supergravity, all AdS solutions come in families labelled by two scaling parameters. One can check that the lower-dimensional cosmological constant and the KK scale depend on these parameters in such a way that no parametric scale separation occurs in this two-dimensional parameter space. On the other hand, in the DGKT vacua, the scaling symmetries are broken due to the presence of the O6-planes. Instead, a \emph{different} scaling symmetry, with corresponding parameter $n$, arises in these solutions due to the vanishing of certain terms in the supergravity equations. This extra parameter is not present in other AdS solutions and, as stated above, it enters the relevant expressions precisely such as to allow for scale separation in the limit $n\to\infty$.
Our arguments thus explain why scale separation does not occur in most AdS solutions in string theory and why the DGKT vacua and their non-supersymmetric cousins behave differently.

Note that the earlier work \cite{Gautason:2015tig} already derived a no-go theorem against scale separation for the case of AdS vacua without O-planes.
However, as we will explain in more detail below, the bound derived there does actually not forbid AdS/KK scale separation in such vacua if the KK scale is small compared to the Planck scale (which should be the case in a controlled flux compactification). It was furthermore pointed out in \cite{Gautason:2015tig} that their argument assumes a specific relation between the KK scale and the internal curvature which need not hold on general manifolds.
The arguments presented in the present paper
do not make these two assumptions
and have, to our knowledge, not been discussed elsewhere.

This paper is organized as follows. In Section \ref{sec:setup}, we establish our conventions, state the type IIA supergravity equations and review the smeared DGKT solutions. In Section \ref{sec:sol}, we construct the exact solutions including the O6-plane backreaction in the large-$n$ regime. In Section \ref{sec:example}, we study the $T^6/{\mathbb{Z}_3^2}$ orientifold as a simple explicit example. We then move on to a discussion of the general 4d scalar potential and its corrections in Section \ref{sec:scalar}. In Section \ref{sec:scale}, we briefly review the strong AdS distance conjecture of \cite{Lust:2019zwm} and the no-go argument of \cite{Gautason:2015tig}. We then discuss scaling symmetries of type II supergravity and their relation to AdS/KK scale separation. We conclude in Section \ref{sec:concl} with a summary of our results and a discussion of future research directions. In Appendix \ref{sec:back}, we illustrate with an example and a counter-example when our general method can be used to construct backreacted solutions in flux compactifications.
\\

\section{Setup}

\label{sec:setup}

In this section, we establish our conventions, state the equations of motion of type IIA supergravity and review the smeared DGKT solutions.

\subsection{Conventions and Equations of Motion}

Our ansatz for the metric in string frame is
\begin{equation}
\d s_{10}^2  = w^2 g_{\mu\nu} \d x^\mu \d x^\nu + g_{mn}\d y^m \d y^n, \label{metric}
\end{equation}
where $g_{\mu\nu}$ is the AdS$_4$ metric with unit radius and $w=w(y)$ denotes the warp factor. In the smeared solutions, $w$ is a constant which determines the AdS scale. The internal metric $g_{mn}$ is Ricci-flat in the smeared solutions but curved if the O6-plane backreaction is included, as we shall see below.

Since we are only interested in vacuum solutions, we will assume that the supergravity fields do not depend on the AdS coordinates $x^\mu$. Furthermore, we only allow NSNS and RR forms with legs along internal directions. An exception is $F_4$, which can be spacetime-filling without breaking the maximal symmetry of the AdS$_4$ factor. In the following, we will write the spacetime-filling piece of $F_4$ in terms of its dual $F_6$, which has only internal legs. Also note that we will work with the usual improved RR field strengths which (locally) satisfy $F_q = \d C_{q-1} - H_3 \w C_{q-3}+ \frac{1}{(q/2)!}F_0 B_2^{q/2}$ for $q=2,4,6$.
Furthermore, we will set $2\pi\sqrt{\alpha^\prime}=1$ in all equations.

The contribution of the localized O6-planes\footnote{One may furthermore allow D6-branes to be present in the compactification, which contribute with the opposite sign.} to the equations of motion is taken into account by terms involving delta distributions.\footnote{As is standard in the literature, we will work with the (physically reasonable) assumption that the localized sources contribute to the supergravity equations with delta-distribution charge densities and energy-momentum, in agreement with the known couplings of an O-plane to the supergravity fields (see, however, \cite{Cordova:2019cvf} for recent doubts). The supergravity solution obtained this way receives string-theory corrections very close to the sources but is valid sufficiently far away from them (see Section \ref{sec:scalar}).} We denote by $\delta(\pi_i)$ the delta distribution with support on the 3-cycle $\pi_i$ wrapped by the $i$th O6-plane
and by $\delta_{i3}$ the corresponding 3-form that integrates to one over the dual cycle $\tilde\pi_i$.
We define $\delta(\pi_i) \equiv \frac{\sqrt{g_{\pi_i}}}{\sqrt{g_6}}\delta^{(3)}(y)$ and $\int_{\tilde\pi_i} \delta_{i3} \equiv \int_{\tilde\pi_i}\d^3 y\, \delta^{(3)}(y) = 1$ in local coordinates such that
\begin{equation}
\int_{\pi_i} \text{vol}_{\pi_i} = \int \text{vol}_{\pi_i} \w \delta_{i3} = \int \d^6 y \sqrt{g_6}\, \delta(\pi_i). \label{delta}
\end{equation}
Here, $g_6\equiv\det(g_{mn})$ and $g_{\pi_i} \equiv \det ((g_{\pi_i})_{\alpha\beta})$ with worldvolume metric $(g_{\pi_i})_{\alpha\beta} = g_{mn} \frac{\partial y^m}{\partial \xi_i^\alpha}\frac{\partial y^n}{\partial \xi_i^\beta}$ and worldvolume coordinates $\xi_i^\alpha$, $\alpha=1,2,3$.

Let us now state the equations of motion for the various fields. The RR and NSNS field equations (in string frame) are\footnote{Note that the Hodge star is defined here with respect to the full 10d metric including the warp factor.}
\begin{align}
0 &= \d \left( \star_{10} F_2 \right) + H_3 \w \star_{10} F_4, \label{eom1} \\
0 &= \d \left( \star_{10} F_4 \right) + H_3 \w \star_{10} F_6, \label{f4} \\
0 &= \d \left( \star_{10} F_6 \right), \\
0 &= \d \left(\tau^2 \star_{10} H_3 \right) + \star_{10} F_2 \w F_0 + \star_{10} F_4 \w F_2 + \star_{10} F_6 \w F_4, \label{eom1b}
\end{align}
where $\tau \equiv \e^{-\phi}$. The Bianchi identities are\footnote{Here, we used that the (downstairs) charge of an O$p$-plane is $-2^{p-5}2\pi$ in our conventions.}
\begin{equation}
\d F_2 = H_3 \w F_0 - 2\sum_i \delta_{i3} \label{f2bianchi}
\end{equation}
and
\begin{equation}
\d F_0 = 0, \quad \d F_4 = H_3 \w F_2, \quad \d F_6 = 0, \quad \d H_3 = 0. \label{bianchis}
\end{equation}
Finally, we state the Einstein and dilaton equations:
\begin{align}
0 &= 12\frac{\tau^2}{w^2} + 12\frac{\tau^2}{w^2} (\partial w)^2 +4\frac{\tau^2}{w} \nabla^2 w + 12\frac{\tau}{w}(\partial w)(\partial \tau) + \tau \nabla^2\tau + (\partial \tau)^2 -\frac{1}{2}\tau^2 |H_3|^2 \nl - \sum_{q=0}^{6} \frac{q-1}{4} |F_q|^2 + \frac{1}{2} \tau \sum_i\delta(\pi_i), \label{eom2c} \\
0 &= -\tau^2 R_{mn} + 4\frac{\tau^2}{w} \nabla_m \partial_n w + \frac{\tau}{w}g_{mn} (\partial w)(\partial \tau)
+ \frac{1}{4} g_{mn} \tau \nabla^2\tau + \frac{1}{4} g_{mn} (\partial \tau)^2 + 2 \tau \nabla_m \partial_n \tau \nl - 2 (\partial_m \tau) (\partial_n\tau) + \frac{1}{2}\tau^2 \left( |H_3|_{mn}^2-\frac{1}{4}g_{mn} |H_3|^2\right)
+\frac{1}{2} \sum_{q=0}^{6} \left( |F_q|_{mn}^2-\frac{q-1}{8}g_{mn} |F_q|^2 \right) \nl + \sum_i \left( \Pi_{i,mn}- \frac{7}{8}g_{mn}\right) \tau \delta(\pi_i), \\
0 &= -8 \nabla^2 \tau - 24\frac{\tau}{w^2} - \frac{32}{w} (\partial w)(\partial \tau) - 24\frac{\tau}{w^2} (\partial w)^2 -16\frac{\tau}{w} \nabla^2 w + 2\tau R_{mn}g^{mn} - \tau |H_3|^2 \nl + 2\sum_i \delta(\pi_i), \label{eom2}
\end{align}
where we used that $R_{\mu\nu} g^{\mu\nu} =- 12$.
The stress-energy of the $i$th O-plane is proportional to the projector
\begin{equation}
\Pi_{i,mn} = -\frac{2}{\sqrt{g_{\pi_i}}} \frac{\delta \sqrt{g_{\pi_i}}}{\delta g^{mn}} = (g_{\pi_i})^{\alpha\beta} \frac{\partial y^l}{\partial \xi_i^\alpha}\frac{\partial y^p}{\partial \xi_i^\beta} g_{ml} g_{np}. \label{proj}
\end{equation}
For example, in the simple case where an O-plane extends along the 4d spacetime and  $y^1,y^2,y^3$, we have $\frac{\partial y^m}{\partial \xi_i^\alpha} = \delta_\alpha^m$. We then obtain $\Pi_{i,mn}=g_{mn}$ along directions parallel to the O-plane and $\Pi_{i,mn}=0$ for transverse directions. In the explicit example discussed in Section \ref{sec:example}, some of the O-planes are diagonal with respect to the $y^m$ coordinates, leading to different expressions for $\Pi_{i,mn}$.
\\

\subsection{The Smeared DGKT Solutions}
\label{sec:setup-smeared}

In the smeared approximation, the equations of motion \eqref{eom1}--\eqref{eom2} are solved under the simplifying assumption that the O-planes are spread out over the whole compact space. Hence, we have to make the replacements
\begin{equation}
\delta(\pi_i) \to j_{\pi_i} = \frac{\mathcal{V}_{\pi_i}}{\mathcal{V}}, \qquad \delta_{i3} \to j_{i3} = \frac{1}{\mathcal{V}_{\tilde\pi_i}} \text{vol}_{\tilde\pi_i} \label{jb0}
\end{equation}
with
\begin{equation}
\mathcal{V}=\int \d^6 y\, \sqrt{g_6}, \quad \mathcal{V}_{\pi_i} = \int_{\pi_i} \d^3y \, \sqrt{g_{\pi_i}}, \quad \mathcal{V}_{\tilde\pi_i} = \int_{\tilde\pi_i} \d^3y \, \sqrt{g_{\tilde\pi_i}}
\end{equation}
such that $\int_{\tilde\pi_i} \delta_{i3} = \int_{\tilde\pi_i} j_{i3} = 1$.

Following \cite{Grimm:2004ua, DeWolfe:2005uu}, we furthermore assume that the field-strength forms are harmonic in the vacuum, that $g_{mn}$ is Ricci-flat and that the warp factor and the dilaton are constant:
\begin{equation}
\d F_q = \d \star_6 F_q = 0, \qquad \d H_3 = \d \star_6 H_3 = 0, \qquad R_{mn} = 0, \qquad \partial_m w = \partial_m \tau = 0. \label{smeared0}
\end{equation}
Under these assumptions, \eqref{f4} can only be satisfied with $H_3\neq 0$ if $F_6=0$.\footnote{Note that this does not imply zero 6-form flux in the 4d superpotential (denoted by $e_0$ in \cite{DeWolfe:2005uu}) since $F_6$ is the \emph{improved} field strength (locally given by $\d C_5 - H_3 \w C_3 + \frac{1}{6}F_0 B_2\w B_2\w B_2$). On the other hand, the flux parameter $e_0$ in $W$ is due to a harmonic piece in the \emph{unimproved} field strength (locally $\d C_5$), which can still be non-zero.}
The remaining non-trivial form-field equations and Bianchi identities are
\begin{align}
0 &= H_3 \w \star_6 F_4, \label{smeared0b} \\
0 &= \star_6 F_2 \w F_0 + \star_6 F_4 \w F_2, \label{smeared0c} \\
0 &= H_3 \w F_0 - 2\sum_i j_{i3}, \label{smeared2} \\
0 &= H_3 \w F_2. \label{smeared2b}
\end{align}

The Einstein and dilaton equations simplify to
\begin{align}
0 &= 12 \frac{\tau^2}{w^2} -\frac{1}{2}\tau^2 |H_3|^2 +\frac{1}{4} |F_0|^2 -\frac{1}{4} |F_2|^2 -\frac{3}{4} |F_4|^2 + \frac{1}{2} \tau \sum_i j_{\pi_i}, \label{smeared1} \\
0 &= \frac{1}{2}\tau^2 \left( |H_3|_{mn}^2-\frac{1}{4}g_{mn} |H_3|^2\right) +\frac{1}{16} g_{mn} |F_0|^2 +\frac{1}{2} \left( |F_2|_{mn}^2-\frac{1}{8}g_{mn} |F_2|^2 \right) \nl +\frac{1}{2} \left( |F_4|_{mn}^2-\frac{3}{8}g_{mn} |F_4|^2 \right) + \sum_i \left( \Pi_{i,mn}- \frac{7}{8}g_{mn}\right) \tau j_{\pi_i}, \\
0 &= -24\frac{\tau}{w^2} -\tau |H_3|^2 + 2\sum_i j_{\pi_i}. \label{smeared3}
\end{align}
All other equations vanish identically with the above ansatz.

In order to solve \eqref{smeared0b}--\eqref{smeared3}, one can now expand the various fields in harmonic forms on the Calabi-Yau and solve the resulting equations for the K\"{a}hler and complex-structure moduli and for the dilaton. Equivalently, the solutions are obtained as extrema of an $F$-term potential in an effective 4d $\mathcal{N}=1$ supergravity description, which was derived in \cite{Grimm:2004ua} by dimensionally reducing type IIA supergravity on a Calabi-Yau orientifold. It was shown in \cite{DeWolfe:2005uu, Acharya:2006ne} that supersymmetric critical points of this potential lift to a 10d smeared solution of the above equations (if in addition the usual tadpole condition for D6 charges is imposed, which corresponds to satisfying the Bianchi identity \eqref{smeared2}). Since the non-supersymmetric solutions found in \cite{Camara:2005dc, Narayan:2010em, Marchesano:2019hfb} are extrema of the same $F$-term potential, we expect that they lift to 10d smeared solutions as well.\footnote{One may wonder whether extrema of the $F$-term potential necessarily satisfy the constraints \eqref{smeared2b} and \eqref{smeared0b}, which do not seem to correspond to any field equation in 4d. Indeed, the former is due to one of the Bianchi identities, which are usually not implied by the equations of motion, and the latter is due to the equation of motion for $C_1$, which is not associated with a modulus since a Calabi-Yau has no harmonic 1-forms. However, one can verify using the results of \cite{DeWolfe:2005uu, Acharya:2006ne, Marchesano:2019hfb} (see also \cite{Behrndt:2004km, Lust:2004ig}) that all 4d solutions (both supersymmetric and non-supersymmetric) satisfy $\star_6 F_4 \sim J$ and either $F_2=0$ or $F_2 \sim J$ (corresponding to $\rho_a \sim \mathcal{K}_a$ and either $\tilde \rho^a =0$ or $\tilde \rho^a \sim t^a$ in the notation of \cite{Marchesano:2019hfb}), where $J$ is the K\"{a}hler form. Since the product of a harmonic form with $J$ is again harmonic, it follows that $H_3 \w F_2$ and $H_3 \w \star_6 F_4$ are harmonic 5-forms, which vanish on a Calabi-Yau.} We refrain from spelling out the various expressions for the moduli here, as we will not need them in the following. It is sufficient for us to know that the smeared solutions satisfy the equations \eqref{smeared0}--\eqref{smeared3}.

We do, however, want to make one last point about the smeared solutions that will be crucial for the remainder of this paper.
In particular, the field equations \eqref{smeared0}--\eqref{smeared3} are invariant under a rescaling of the fields with a parameter $n$ such that
\begin{align}
& F_4 \sim n, \quad F_2 \sim n^{1/2}, \quad F_0, H_3 \sim n^0, \quad \tau \sim n^{3/4}, \quad w \sim n^{3/4}, \quad g_{mn} \sim n^{1/2}. \label{scalings}
\end{align}
The smeared solutions therefore have a free parameter $n$, which is proportional to the amount of $F_4$ flux. It was already observed in \cite{DeWolfe:2005uu} that the large-$n$ limit corresponds to the limit of large volume and small $g_s$.
In this limit, $\alpha^\prime$ and loop corrections to the supergravity equations are parametrically suppressed such that the supergravity solutions are well-controlled. We will see below that corrections due to the O-plane backreaction are parametrically controlled as well in this limit.

In the following, it will be convenient to make the above scaling symmetry manifest by writing
\begin{align}
 F_4 &= F^{(0)}_4 n, &
F_2 &= F^{(0)}_2 n^{1/2}, &
F_0 &= F^{(0)}_0, &
H_3 &= H^{(0)}_3, \notag \\
\tau &= \tau^{(0)}n^{3/4}, &
w &= w^{(0)}n^{3/4}, &
g_{mn} &= g_{mn}^{(0)} n^{1/2}. &&
\end{align}
The $n$-scaling thus cancels out in the field equations \eqref{smeared0}--\eqref{smeared3} when expressed in terms of the quantities labelled by ``(0)''. The convenience of this notation will become clear in the next section. There, we will show that the exact solutions, which include the full O-plane backreaction, can in fact be written in terms of a $1/n$-expansion in which the above smeared field expressions turn out to be the leading-order terms.
\\

\section{The Exact Solutions}

\label{sec:sol}

We are now ready to discuss the solutions including the backreaction of the O6-planes. Since supergravity is not reliable at small volumes and large $g_s$, we will only be interested in the large-$n$ regime.
We expect that the full solution in this regime is given by the smeared solution plus a correction which is subleading in $n$ and captures the backreaction. As we will see momentarily, this intuition is indeed correct.

We make the following ansatz for the fields in the large-$n$ regime:
\begin{align}
F_6 &= 0 + F^{(1)}_6 n + \mathcal{O}(n^{1/2}), \label{ansatz0} \\
F_4 &= F^{(0)}_4 n + F^{(1)}_4 n^{1/2} + \mathcal{O}(n^0), \\
F_2 &= F_2^{(0)} n^{1/2} + F^{(1)}_2 n^{0} + \mathcal{O}(n^{-1/2}), \label{ansatzf} \\
H_3 &= H^{(0)}_3 n^0 + H^{(1)}_3 n^{-1/2} + \mathcal{O}(n^{-1}), \label{ansatzh} \\
\tau &= \tau^{(0)}n^{3/4} + \tau^{(1)} n^{-1/4} + \mathcal{O}(n^{-5/4}), \label{ansatzw} \\
w &= w^{(0)}n^{3/4} + w^{(1)}n^{-1/4} + \mathcal{O}(n^{-5/4}), \label{sol} \\
g_{mn} &= g_{mn}^{(0)} n^{1/2} + g_{mn}^{(1)}n^{-1/2} + \mathcal{O}(n^{-3/2}). \label{ansatz}
\end{align}
Here, the quantities with a superscript ``(0)'' are assumed to equal the smeared solution, which satisfies \eqref{smeared0}--\eqref{smeared3}. Note that we do not expand the Romans mass $F_0=F_0^{(0)}$ since it is a quantized constant and cannot receive any $1/n$ corrections.

We now substitute the above ansatz into the equations of motion \eqref{eom1}--\eqref{eom2} and expand in $1/n$. The $F_2$ Bianchi identity thus yields
\begin{align}
\d \left( F_2^{(0)} n^{1/2} + F_2^{(1)} + \ldots \right) &= \left( H_3^{(0)} + H^{(1)}_3 n^{-1/2} + \ldots \right) \w F_0^{(0)}
- 2\sum_i \delta_{i3}.
\end{align}
Note that the delta form on the right does not receive $1/n$ corrections since it is independent of the metric and the other fields according to its definition above \eqref{delta}.
Since $\d F_2^{(0)} = 0$ by \eqref{smeared0}, we find at leading order in $1/n$:
\begin{equation}
\d F_2^{(1)} = H_3^{(0)}\w F_0^{(0)}- 2\sum_i \delta_{i3}. \label{loc1}
\end{equation}
The remaining Bianchi identities, again at leading order in $1/n$, yield
\begin{align}
\d F_6^{(1)} = 0, \quad \d F_4^{(1)} = H_3^{(0)}\w F_2^{(0)} = 0, \quad \d H_3^{(1)} = 0, \label{loc2}
\end{align}
which is satisfied for any closed $F_6^{(1)}$, $F_4^{(1)}$ and $H_3^{(1)}$.
Here, we used that \eqref{smeared0} and \eqref{smeared2b} hold for the smeared solution.
One also verifies using \eqref{smeared0}--\eqref{smeared0c} that the form field equations \eqref{eom1}--\eqref{eom1b} are satisfied at leading order in $1/n$ if
\begin{align}
\d \star_6^{(0)}\! F_6^{(1)} = \d \star_6^{(0)}\! F_4^{(1)} = \d \star_6^{(0)}\! F_2^{(1)} = \d \star_6^{(0)}\! H_3^{(1)} = 0. \label{loc2b}
\end{align}

Substituting our ansatz into \eqref{eom2c}--\eqref{eom2}, we furthermore find the leading-order Einstein and dilaton equations:
\begin{align}
0 &= 12\frac{\tau^{(0)2}}{w^{(0)2}} + 4 \frac{(\tau^{(0)})^2}{w^{(0)}} \nabla^2 w^{(1)} + \tau^{(0)} \nabla^2\tau^{(1)} -\frac{1}{2}(\tau^{(0)})^2 |H_3^{(0)}|^2
-\sum_{q=0}^{4} \frac{q-1}{4}  |F_q^{(0)}|^2 \nl + \frac{1}{2} \tau^{(0)} \sum_i\delta(\pi_i), \label{loc4} \\
0 &= -\tau^{(0)2} R^{(1)}_{mn} + 4\frac{\tau^{(0)2}}{w^{(0)}} \nabla_m \partial_n w^{(1)} + \frac{1}{4} g_{mn}^{(0)} \tau^{(0)} \nabla^2\tau^{(1)} + 2 \tau^{(0)} \nabla_m \partial_n \tau^{(1)} \nl + \frac{1}{2}(\tau^{(0)})^2 \left( |H_3^{(0)}|_{mn}^2-\frac{1}{4}g^{(0)}_{mn} |H^{(0)}_3|^2\right)
+\frac{1}{2} \sum_{q=0}^{4} \left( |F^{(0)}_q|_{mn}^2-\frac{q-1}{8}g_{mn}^{(0)} |F^{(0)}_q|^2 \right) \nl + \sum_i \left( \Pi^{(0)}_{i,mn}- \frac{7}{8}g^{(0)}_{mn}\right) \tau^{(0)} \delta(\pi_i), \label{loc5} \\
0 &= - 8 \nabla^2 \tau^{(1)}- 24\frac{\tau^{(0)}}{w^{(0)2}} - 16 \frac{(\tau^{(0)})}{w^{(0)}} \nabla^2 w^{(1)} + 2\tau^{(0)} R_{mn}^{(1)}g^{(0)mn} -\tau^{(0)} |H_3^{(0)}|^2  \nl + 2\sum_i \delta(\pi_i) \label{loc3}
\end{align}
with
\begin{align}
R^{(1)}_{mn} &= - \frac{1}{2} g^{(0)rs} \nabla_m \nabla_n g^{(1)}_{rs}+\frac{1}{2} g^{(0)rs} \left( \nabla_s \nabla_m g^{(1)}_{rn}+\nabla_s \nabla_n g^{(1)}_{rm}\right) -\frac{1}{2} \nabla^2 g_{mn}^{(1)}. \label{r}
\end{align}
Here and in the following, we do not display a superscript ``(0)'' on covariant derivatives and source terms to avoid cluttering the equations with too many indices. The reader should keep in mind that from now on we denote by $\nabla_m$ the covariant derivative adapted to $g_{mn}^{(0)}$ and that metric determinants implicit in $j_{\pi_i}$, $\delta(\pi_i)$ should be taken with respect to $g_{mn}^{(0)}$ as well (i.e., $\nabla_m \equiv \nabla_m^{(0)}$, $j_{\pi_i} \equiv j_{\pi_i}^{(0)}$, $\delta(\pi_i) \equiv \delta^{(0)}(\pi_i)$).

In order to simplify the above equations, we now use the smeared equations \eqref{smeared2}, \eqref{smeared1}--\eqref{smeared3} and substitute the terms labelled by ``(0)'' by the smeared sources $j_{\pi_i}$. Eqs.~\eqref{loc1}, \eqref{loc4}--\eqref{loc3} can thus be written as
\begin{align}
\d F_2^{(1)} &=  2\sum_i (j_{i3}-\delta_{i3}), \label{final-eom1} \\
\nabla^2 \tau^{(1)} &= -\frac{3}{2}\sum_i \left( j_{\pi_i}- \delta(\pi_i) \right), \label{final-eom2} \\
\nabla^2 w^{(1)} &= \frac{1}{2} \frac{w^{(0)}}{\tau^{(0)}}  \sum_i \left( j_{\pi_i}- \delta(\pi_i) \right), \label{final-eom3} \\
\tau^{(0)} R^{(1)}_{mn} - 4 \frac{\tau^{(0)}}{w^{(0)}}\nabla_m \partial_n w^{(1)} - 2 \nabla_m \partial_n \tau^{(1)} &= \sum_i \left( \frac{1}{2}g^{(0)}_{mn}-\Pi^{(0)}_{i,mn}\right)  \left( j_{\pi_i}- \delta(\pi_i) \right). \label{final-eom4}
\end{align}
Eqs.~\eqref{final-eom1}--\eqref{final-eom4} are one of the main results of this paper. Together with \eqref{loc2} and \eqref{loc2b}, they fully determine the O-plane backreaction in the large-volume and small-$g_s$ regime for a given orientifold compactification.

Indeed, we observe that the corrections labelled by ``(1)'' precisely account for the difference $j_{\pi_i}- \delta(\pi_i)$ between smeared and localized sources as expected.
We have thus shown that, at large $n$, there is a one-to-one correspondence between a smeared solution on a particular Calabi-Yau orientifold $X_0$ and the exact (i.e., fully backreacted) solution on a curved orientifold $X$, where the two solutions differ by the terms labelled by ``(1)''. Crucially, these terms appear at leading order in the equations of motion
but at subleading order in the fields themselves (cf.~\eqref{ansatz0}--\eqref{ansatz}). The exact solution is therefore equal to the smeared one plus a small correction.

To be precise, the correction is small \emph{almost} everywhere on the compact space, except at distances $\lesssim \mathcal{O}(1)$ (in string units) very close to the O-planes where non-linear backreaction effects and string corrections become relevant.
We will come back to this point in Section \ref{sec:scalar}, where we will analyze the validity of our large-$n$ expansion in more detail. Since our expansion breaks down near the source positions, the above equations are not valid there, and one may wonder what the significance of the source terms is on their right-hand sides. The assumption here is that the delta distributions source the correct long-distance behavior of the fields compatible with the charges and the tension of the O-planes.\footnote{Alternatively, one may solve the equations without any source terms and instead impose appropriate boundary conditions for the fields at some distance $r=r_0$ where our expansion is still valid (such that the charges and the tension of the sources inside the small-$r$ regions are reproduced), which should lead to the same result.} The equations should therefore be read as determining the linearized supergravity solutions sufficiently far away from the O-planes, and it is understood that they cease to be valid at small distances.

Let us also stress that our solutions at this point have the same level of explicitness as the well-known GKP vacua in type IIB \cite{Giddings:2001yu, Dasgupta:1999ss}, where the supergravity equations are solved up to a Poisson equation with O-plane/D-brane sources (whose explicit solution depends on the considered orientifold). Similarly, the solutions in our case are given in terms of the simple equations \eqref{final-eom1}--\eqref{final-eom4}. In Section \ref{sec:example}, we will solve these equations explicitly on a specific toroidal orientifold.

We close this section with a few comments on how our solution relates to various statements and conjectures about backreaction/warping in the literature:
\begin{itemize}
\item It was argued in \cite{Douglas:2010rt} that backreaction/warping corrections have to be of the order of the fluxes in the 10d equations of motion and are therefore never negligible (see also \cite{Blaback:2010sj, Junghans:2013xza}). This essentially follows from the fact that a smeared source can cancel flux terms in equations such as \eqref{f2bianchi} everywhere on the compact manifold while a localized source cannot do this. Therefore, the localized solution requires corrections of the order of the flux terms which account for the missing energy density away from the source loci. Indeed, the backreaction terms in our solution
appear at leading order in the equations of motion as expected from the discussion in \cite{Douglas:2010rt}.
\item It is often claimed in the flux compactification literature that backreaction/warping becomes negligible (i.e., smearing becomes a good approximation) at large volumes and small $g_s$ (see, e.g., \cite{Acharya:2006ne}). This is again true in our solutions, as the corrections to the smeared field expressions become small at large $n$, cf.~\eqref{ansatz0}--\eqref{ansatz}. As we will discuss more explicitly in Section \ref{sec:scalar}, this property also ensures that the corrections do not significantly alter the 4d scalar potential. One may think that this statement contradicts our previous claim in bullet point 1. However, as stated above, it is crucial here to distinguish between backreaction corrections to the 10d equations of motion (which are leading) and backreaction corrections to the fields themselves (which are subleading).

\item In \cite{Saracco:2012wc}, the authors studied the local solution near an O6-plane in a supersymmetric AdS background in the presence of Romans mass. A numerical analysis then revealed the existence of regular solutions (i.e., neither the curvature nor the dilaton blows up close to the O-plane). This was interpreted as a possible resolution mechanism for O6-plane singularities in massive type IIA string theory where no M-theory lift is available. The solution in \cite{Saracco:2012wc} assumes exact $H_3$ and no O6-plane intersections so that it does not directly relate to the DGKT case.
Nevertheless, one may speculate that a similar mechanism is realized there as well.

As we will see more explicitly in Sections \ref{sec:example} and \ref{sec:scalar}, our solution does not resemble the one of \cite{Saracco:2012wc} in the region close to the O-planes but rather exhibits metric and dilaton singularities near the O-planes. However, as stated before, our large-$n$ expansion cannot be trusted at small distances from the O6-planes (cf.~again Section \ref{sec:scalar}). We therefore do not know whether non-linear backreaction effects conspire in such a way at the O6-planes that the local field behavior is similar to the one in \cite{Saracco:2012wc}.
Note, however, that our solution is more reminiscent of the GKP solution and its T-duals in the sense that the backreacted and smeared field expressions differ by terms satisfying Poisson equations on the (un-backreacted) Calabi-Yau manifold \cite{Blaback:2010sj}. The fields are therefore given by Green functions of the corresponding Laplacians, with the usual divergences associated to Dirac-delta sources. In GKP, this behavior persists including the full non-linear backreaction. In particular, the fields there agree with the known flat-space behavior of the O-planes in their vicinity, which implies that the curvature and the dilaton blow up.

\item Another work studying the
resolution of the O6-plane singularities is \cite{Gautason:2015tig}.
There, it was argued
that AdS solutions of type II (or 11d) supergravity cannot have scale separation unless they have explicit O-plane sources or large integrated dilaton gradients.\footnote{``Large'' here means $\int \d^6 y \sqrt{g_6}\, w^4(\partial \tau)^2 \gg \sum_q \int \d^6 y \sqrt{g_6}\, w^4 |F_q|^2$ in string frame.} Since the regular solution found in \cite{Saracco:2012wc} has no explicit O-planes and an almost constant dilaton, \cite{Gautason:2015tig} concluded that it cannot appear in vacua with scale separation such as the DGKT vacua.
However, as we will discuss in more detail in Section \ref{sec:scale}, the argument of \cite{Gautason:2015tig} is based on two assumptions about the KK scale which do not hold in general (and are indeed violated in the DGKT vacua).
We therefore believe that
the argument does not imply that a solution with resolved O-plane singularities would require large dilaton gradients. Consequently, a resolution mechanism of the type discussed in \cite{Saracco:2012wc} is not ruled out in the DGKT setup. Nevertheless, we stress again that whether such a regular solution exists is currently not known, and it would be very interesting to understand this better.
\\

\end{itemize}

\section{A Simple Example}
\label{sec:example}

In this section, we work out the backreaction for the case where the internal space is an orientifold of the $T^6/{\mathbb{Z}_3^2}$ orbifold \cite{Dixon:1985jw, Strominger:1985ku, Blumenhagen:1999ev}. This simple model was already studied in \cite{DeWolfe:2005uu} as their prime example.
In Section \ref{sec:ex-symm}, we will review the discrete symmetries of the model, the torus identifications and the orientifold involution. We will then discuss our ansatz for the internal metric $g_{mn}$ in Section \ref{sec:ex-ansatz}. In Section \ref{sec:ex-sol}, we will explicitly solve \eqref{final-eom1}--\eqref{final-eom4} on this background.\footnote{It will be convenient in this section to study the model from the point of view of the covering torus. This introduces image O-planes and effectively doubles the charge of each O-plane compared to the downstairs equations stated before.}

\subsection{Symmetries}
\label{sec:ex-symm}

In the following, it will be convenient to work with complex coordinates $z_a$ ($a=1,2,3$), which are related to the real coordinates $y^m$ ($m=1,\ldots,6$) by
\begin{equation}
z_1 = y^1+iy^2, \quad z_2 = y^3+iy^4, \quad z_3 = y^5+iy^6. \label{coord}
\end{equation}
On the $T^6$, the coordinates are periodically identified such that
\begin{align}
& z_a \sim z_a + 1 \sim z_a+\alpha \label{torus}
\end{align}
with $\alpha= \e^{i\pi/3}$.
The torus has two $\mathbb{Z}_3$ symmetries $T$ and $Q$, which act as
\begin{align}
T: z_a \to \alpha^2 z_a, \qquad Q: z_a \to \alpha^{2a} z_a + \frac{1+\alpha}{3}. \label{qt}
\end{align}
Modding out by these two symmetries, we obtain a $T^6/{\mathbb{Z}_3^2}$ orbifold. The model we consider is an orientifold of this orbifold, where the orientifold involution acts on the $z_a$ as
\begin{equation}
\sigma: z_a \to -z_a^*. \label{orientifold}
\end{equation}

\begin{figure}[t]
\centering
\includegraphics[trim = 0mm 20mm 0mm 20mm, clip, width=0.4\textwidth]{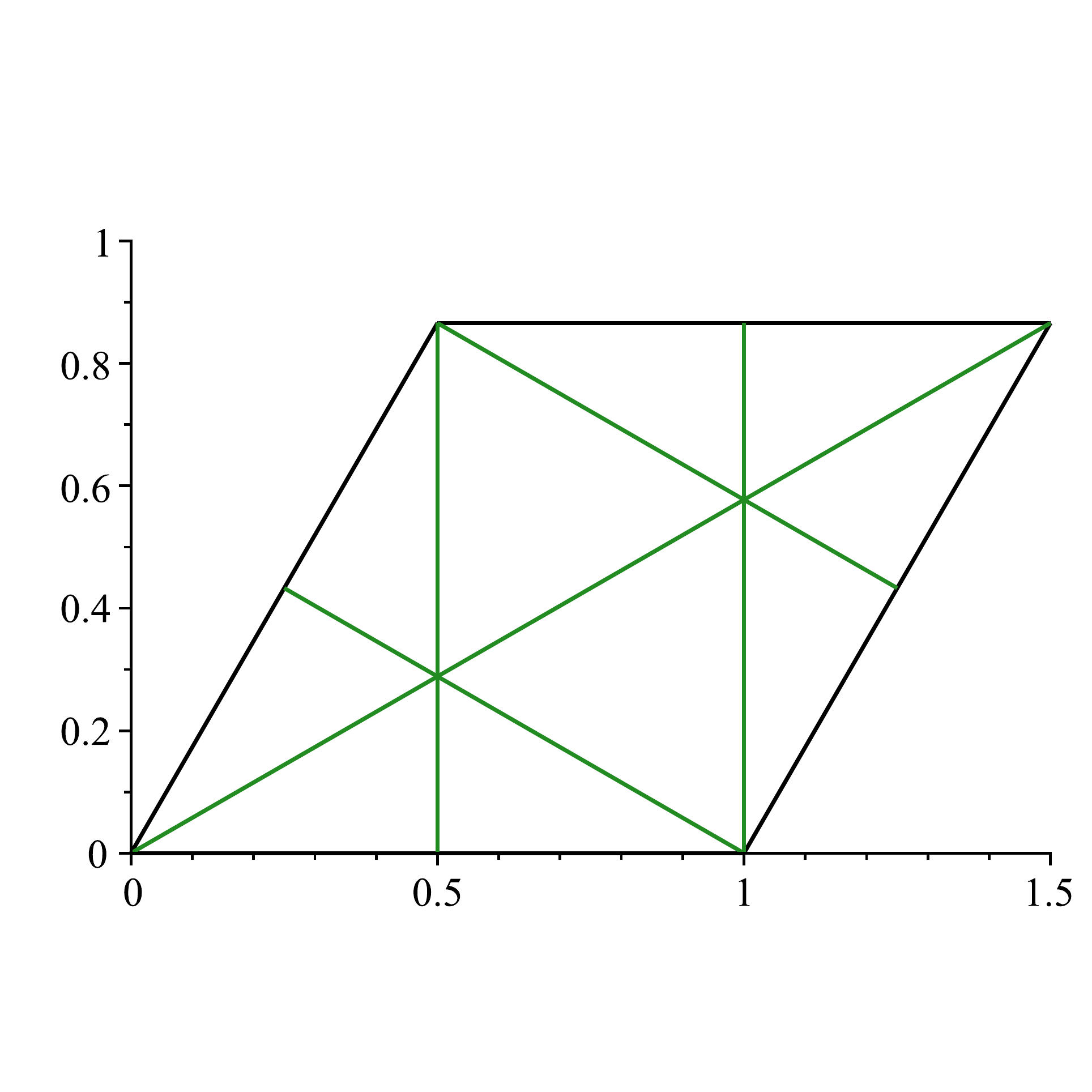}
\put(-180,115){$\scriptstyle{y^2}$}
\put(-25,5){$\scriptstyle{y^1}$}
\caption{Fundamental domain of the 2-torus parametrized by $y^1$, $y^2$ with O-plane images under actions of $Q$ and $T$. \label{fund}}
\end{figure}

Let us now analyze the fixed points $z_a=\hat z_a$ of the orientifold involution, which satisfy $\sigma(\hat z_a)=\hat z_a$. Using \eqref{orientifold} together with \eqref{torus}, we find that fixed points satisfy $\text{Re}(\hat z_a) = \mathbb{Z}/2$. This yields an O6-plane localized at
\begin{equation}
y^1 \in \left\{ \frac{1}{2},1\right\}, \quad y^3 \in \left\{ \frac{1}{2},1\right\}, \quad y^5 \in \left\{ \frac{1}{2},1\right\}, \label{18}
\end{equation}
up to periodic identifications. Here, we only wrote down loci which intersect the fundamental domain of the torus. Note that the identifications \eqref{torus} are such that these eight loci actually describe a single O6-plane winding twice around each 2-torus.

Since the orientifolding is performed on the orbifold $T^6/\mathbb{Z}_3^2$, we also have to take into account all image O-planes under the actions of $Q$ and $T$ from the point of view of the covering torus.
In particular, if $\hat z_a$ is a fixed point, also the points $Q(\hat z_a)$, $T(\hat z_a)$, $Q^2(\hat z_a)$, $T^2(\hat z_a)$, $QT(\hat z_a)$, $QT^2(\hat z_a)$, $Q^2T(\hat z_a)$ and $Q^2T^2(\hat z_a)$ must be fixed points.\footnote{Note that higher powers or different orderings of $Q$ and $T$ do not lead to further images because $Q^3(\hat z_a)\sim T^3(\hat z_a) = \hat z_a$ and $QT(\hat z_a) \sim TQ(\hat z_a)$ on the torus.}

For example, consider the points $\tilde z_a = Q^2(\hat z_a)$. Since $Q^2(\hat z_a) \sim Q^{-1}(\hat z_a)$, they satisfy
\begin{equation}
\sigma(Q(\tilde z_a)) = Q(\tilde z_a)
\end{equation}
up to periodic identifications. This yields the condition
\begin{equation}
\text{Re}\left(\alpha^{2a}\tilde z_a+\frac{1+\alpha}{3}\right) = \frac{\mathbb{Z}}{2}.
\end{equation}
We thus have an O-plane image at
\begin{equation}
y^1 +\sqrt{3}y^2 \in \left\{ 1,2\right\}, \quad y^3-\sqrt{3}y^4 = 0, \quad y^5 \in \left\{ \frac{1}{2},1\right\}. \label{9}
\end{equation}

\begin{table}[t]\renewcommand{\arraystretch}{1.1}\setlength{\tabcolsep}{6pt}
\hspace{-0.65cm}\begin{tabular}{|l|l| l l l|}
\hline 
$\pi_i$ & Fixed point equation & \multicolumn{3}{l|}{O-plane position}  \\
\hline 
 $\pi_0$ & $\sigma(z_a) = z_a$ & $y^1 \in \left\{\frac{1}{2},1\right\}$ & $ y^3 \in \left\{\frac{1}{2},1\right\}$ & $ y^5 \in \left\{\frac{1}{2},1\right\}$ \\
 $\pi_+$ & $\sigma(T(z_a)) = T(z_a)$ & $y^1+\sqrt{3}y^2\in \left\{1,2\right\}$ & $y^3+\sqrt{3}y^4\in \left\{1,2\right\}$ & $y^5+\sqrt{3}y^6\in \left\{1,2\right\}$ \\
 $\pi_-$ & $\sigma(T^2(z_a)) = T^2(z_a)$ & $y^1-\sqrt{3}y^2=0$ & $y^3 -\sqrt{3}y^4=0$ & $y^5-\sqrt{3}y^6=0$ \\
 $\pi_{23}$ & $\sigma(QT^2(z_a)) = QT^2(z_a)$ & $y^1 \in \left\{\frac{1}{2},1\right\}$ & $y^3+\sqrt{3}y^4\in \left\{1,2\right\}$ & $y^5-\sqrt{3}y^6=0$ \\
 $\pi_{32}$ & $\sigma(Q^2T(z_a)) = Q^2T(z_a)$ & $y^1 \in \left\{\frac{1}{2},1\right\}$ & $y^3-\sqrt{3}y^4=0$ & $y^5+\sqrt{3}y^6\in \left\{1,2\right\}$ \\
 $\pi_{31}$ & $\sigma(QT(z_a)) = QT(z_a)$ & $y^1-\sqrt{3}y^2=0$ & $y^3 \in \left\{\frac{1}{2},1\right\}$ & $y^5+\sqrt{3}y^6\in \left\{1,2\right\}$ \\
 $\pi_{13}$ & $\sigma(Q^2T^2(z_a)) = Q^2T^2(z_a)$ & $y^1+\sqrt{3}y^2\in \left\{1,2\right\}$ & $y^3 \in \left\{\frac{1}{2},1\right\}$ & $y^5-\sqrt{3}y^6=0$ \\
 $\pi_{12}$ & $\sigma(Q(z_a)) = Q(z_a)$ & $y^1+\sqrt{3}y^2\in \left\{1,2\right\}$ & $y^3-\sqrt{3}y^4=0$ & $y^5\in \left\{\frac{1}{2},1\right\}$ \\
 $\pi_{21}$ & $\sigma(Q^2(z_a)) = Q^2(z_a)$ & $y^1-\sqrt{3}y^2=0$ & $y^3+\sqrt{3}y^4\in \left\{1,2\right\}$ & $y^5\in \left\{\frac{1}{2},1\right\}$ \\
\hline 
\end{tabular}
\caption{Image O-planes on the covering torus.}
\label{tab-pos}
\end{table}

Analogously, we can compute the image fixed points obtained by further actions of $Q$'s and $T$'s. The full set of O-planes (up to periodic identifications) is given in Table \ref{tab-pos}, see also Fig.~\ref{fund}.
We thus have O-plane images localized in 9 different directions on the covering torus (plus an infinite number of images due to the periodic identifications),
where we denote the corresponding cycles by $\pi_0$, $\pi_\pm$, $\pi_{ab}$.
Note that, on the quotient space $T^6/{\mathbb{Z}_3^2}$, all O-plane images are identified with the eight loci in $\pi_0$. As stated above, these correspond to a single O-plane.

Finally note that a smooth Calabi-Yau is obtained from the orbifold by blowing up 9 singular orbifold points. The volumes of the associated blow-up cycles need to be large in order to control string corrections, which was argued to be possible in the smeared case in \cite{DeWolfe:2005uu}. For simplicity, we will restrict our following analysis to the orbifold limit and leave a detailed study of backreaction corrections in smooth Calabi-Yau models for future work. Note, however, that our general solution in Section \ref{sec:sol} is valid for any Calabi-Yau.
\\

\begin{table}[t]\renewcommand{\arraystretch}{1.2}\setlength{\tabcolsep}{10pt}
\centering
\begin{tabular}{|c|c c c|}
\hline 
$\pi_i$ & \multicolumn{3}{c|}{Projectors}  \\
\hline 
$\pi_0$ & $\Pi^{(0)}_{11} = 0$ & $\Pi^{(0)}_{33} = 0$ & $\Pi^{(0)}_{55} = 0$ \\
 & $\Pi^{(0)}_{22} = v_1^{(0)2}$ & $\Pi^{(0)}_{44} = v_2^{(0)2}$ & $\Pi^{(0)}_{66} = v_3^{(0)2}$ \\
 & $\Pi^{(0)}_{12} = 0$ & $\Pi^{(0)}_{34} = 0$ & $\Pi^{(0)}_{56} = 0$ \\
\hline 
$\pi_\pm$ & $\Pi^{(0)}_{11} = \frac{3}{4} v_1^{(0)2}$ & $\Pi^{(0)}_{33} = \frac{3}{4}v_2^{(0)2}$ & $\Pi^{(0)}_{55} = \frac{3}{4}v_3^{(0)2}$ \\
 & $\Pi^{(0)}_{22} = \frac{1}{4} v_1^{(0)2}$ & $\Pi^{(0)}_{44} = \frac{1}{4}v_2^{(0)2}$ & $\Pi^{(0)}_{66} = \frac{1}{4}v_3^{(0)2}$  \\
 & $\Pi^{(0)}_{12} = \mp \frac{\sqrt{3}}{4} v_1^{(0)2}$ & $\Pi^{(0)}_{34} = \mp \frac{\sqrt{3}}{4} v_2^{(0)2}$  & $\Pi^{(0)}_{56} = \mp \frac{\sqrt{3}}{4} v_3^{(0)2}$  \\
\hline 
$\pi_{23}$, $\pi_{32}$ & $\Pi^{(0)}_{11} = 0$ & $\Pi^{(0)}_{33} = \frac{3}{4}v_2^{(0)2}$ & $\Pi^{(0)}_{55} = \frac{3}{4}v_3^{(0)2}$\\
 &  $\Pi^{(0)}_{22} = v_1^{(0)2}$ &  $\Pi^{(0)}_{44} = \frac{1}{4}v_2^{(0)2}$ & $\Pi^{(0)}_{66} = \frac{1}{4}v_3^{(0)2}$ \\
 & $\Pi^{(0)}_{12} = 0$ & $\Pi^{(0)}_{34} = \mp \frac{\sqrt{3}}{4} v_2^{(0)2}$ & $\Pi^{(0)}_{56} = \pm \frac{\sqrt{3}}{4}v_3^{(0)2}$  \\
\hline 
$\pi_{31}$, $\pi_{13}$ & $\Pi^{(0)}_{11} = \frac{3}{4} v_1^{(0)2}$  &  $\Pi^{(0)}_{33} = 0$  & $\Pi^{(0)}_{55} = \frac{3}{4}v_3^{(0)2}$  \\
 & $\Pi^{(0)}_{22} =  \frac{1}{4}v_1^{(0)2}$ & $\Pi^{(0)}_{44} =v_2^{(0)2}$ & $\Pi^{(0)}_{66} = \frac{1}{4}v_3^{(0)2}$ \\
 & $\Pi^{(0)}_{12} = \pm \frac{\sqrt{3}}{4} v_1^{(0)2}$ & $\Pi^{(0)}_{34} = 0$ & $\Pi^{(0)}_{56} = \mp \frac{\sqrt{3}}{4}v_3^{(0)2}$ \\
\hline
$\pi_{12}$, $\pi_{21}$ & $\Pi^{(0)}_{11} = \frac{3}{4} v_1^{(0)2}$  & $\Pi^{(0)}_{33} = \frac{3}{4}v_2^{(0)2}$ &  $\Pi^{(0)}_{55} =0$ \\
 & $\Pi^{(0)}_{22} =  \frac{1}{4} v_1^{(0)2}$ &  $\Pi^{(0)}_{44} = \frac{1}{4}v_2^{(0)2}$ & $\Pi^{(0)}_{66} = v_3^{(0)2}$ \\
 & $\Pi^{(0)}_{12} = \mp \frac{\sqrt{3}}{4} v_1^{(0)2}$ & $\Pi^{(0)}_{34} = \pm \frac{\sqrt{3}}{4} v_2^{(0)2}$ & $\Pi^{(0)}_{56} = 0$  \\
\hline
\end{tabular} 
\caption{Projectors for image O-planes.}
\label{tab-proj}
\end{table}

\subsection{Metric Ansatz}
\label{sec:ex-ansatz}

Our ansatz for the internal metric in complex coordinates is
\begin{align}
\d s_6^2 &= \sum_{a=1}^3\left(v_{a}^2 \d z_a \d z_a^* + u_{a} \d z_a \d z_a + u_{a}^* \d z_a^* \d z_a^*\right). \label{complex}
\end{align}
Here, $v_a=v_a(z_b,z_b^*)$ is a real function and $u_a=u_a(z_b,z_b^*)$ is a complex function of the internal coordinates, which will both be determined below.
In order for \eqref{complex} to be invariant under the symmetries \eqref{qt}, we require
\begin{align}
v_a\left(z_b, z_b^*\right) &= v_a\left(\alpha^2 z_b, \alpha^{-2} z_b^*\right) = v_a\left(\alpha^{2b} z_b + \frac{1+\alpha}{3}, \alpha^{-2b} z_b^*+ \frac{1+\alpha^{-1}}{3}\right), \\ u_a\left(z_b, z_b^*\right) &= \alpha^{4} u_a\left(\alpha^2 z_b, \alpha^{-2} z_b^*\right) = \alpha^{4a} u_a\left(\alpha^{2b} z_b+ \frac{1+\alpha}{3}, \alpha^{-2b} z_b^*+ \frac{1+\alpha^{-1}}{3}\right) \label{ua}
\end{align}
with $\alpha = \e^{i\pi/3}$.

Comparing \eqref{complex} with \eqref{metric} and \eqref{coord}, we find that $g_{mn}$ is related to $v_a$ and $u_a$ as follows:
\begin{align}
&g_{11} = v_1^2 + 2\text{Re}\,u_1, && g_{22} = v_1^2 - 2\text{Re}\,u_1, && g_{12} = - 2\text{Im}\,u_1, \label{real1} \\
&g_{33} = v_2^2 + 2\text{Re}\,u_2, && g_{44} = v_2^2 - 2\text{Re}\,u_2, && g_{34} = - 2\text{Im}\,u_2, \\
&g_{55} = v_3^2 + 2\text{Re}\,u_3, && g_{66} = v_3^2 - 2\text{Re}\,u_3, && g_{56} = - 2\text{Im}\,u_3 \label{real3}
\end{align}
and $g_{mn}=0$ otherwise. Hence, $g_{mn}$ is block-diagonal in our ansatz.

According to our large-$n$ expansion \eqref{ansatz}, we furthermore write
\begin{equation}
v_a = v_a^{(0)} n^{1/4} + v_a^{(1)}n^{-3/4} + \mathcal{O}(n^{-7/4}), \qquad u_a = u_a^{(0)} n^{1/2} + u_a^{(1)}n^{-1/2}+ \mathcal{O}(n^{-3/2}), \label{uvexp}
\end{equation}
where the superscript ``(0)'' denotes the smeared solution as usual. Recall that $g_{mn}=g_{mn}^{(0)}n^{1/2}$ is Ricci-flat in the smeared solution (see Section \ref{sec:setup-smeared}). In the toroidal case discussed here, it is simply constant, which implies that $v_a^{(0)}$ and $u_a^{(0)}$ are constants. Consistency with \eqref{ua} further restricts this to
\begin{equation}
v_a^{(0)} = \text{const.}, \qquad u_a^{(0)} = 0.
\end{equation}
In the smeared solution, the metric is therefore
\begin{equation}
g_{11}^{(0)}=g_{22}^{(0)}=v_1^{(0)2}, \quad g_{33}^{(0)}=g_{44}^{(0)}=v_2^{(0)2}, \quad g_{55}^{(0)}=g_{66}^{(0)}=v_3^{(0)2} \label{sm-metric}
\end{equation}
and $g_{mn}^{(0)}=0$ for $m\neq n$.

In the next section, we will also need to know the projectors $\Pi^{(0)}_{i,mn}$ which determine the O-plane stress-energy at leading order in $1/n$ (cf.~\eqref{final-eom4}). Using the above metric ansatz, they can be computed from the definition in \eqref{proj}. To this end, we require the relation between the torus coordinates $y^m$ and the worldvolume coordinates $\xi^\alpha$, which can be inferred from Table \ref{tab-pos}. The result for the various O-plane images is summarized in Table \ref{tab-proj}.\footnote{As an example, consider an O-plane localized at $y^1 =1$, $y^3+\sqrt{3}y^4=1$, $y^5-\sqrt{3}y^6=0$. The worldvolume coordinates are then $x^\mu$ for $\mu=0,1,2,3$ and $\xi^1=y^2$, $\xi^2 = \frac{1}{2}\left(\sqrt{3}y^3-y^4\right)$, $\xi^3 = \frac{1}{2}\left(\sqrt{3}y^5+y^6\right)$ and the transverse coordinates are $\chi^1 =y^1$, $\chi^2 = \frac{1}{2}\left(y^3+\sqrt{3}y^4\right)$, $\chi^3 = \frac{1}{2}\left(y^5-\sqrt{3}y^6\right)$. We can now solve for $y^m$ and compute $\partial y^m/\partial \xi^\alpha$ and $(g_{\pi})_{\alpha\beta}$. Using this and \eqref{sm-metric} in \eqref{proj}, we obtain the entry $\pi_{23}$ in Table \ref{tab-proj}.}

Finally, we will use that, in the smeared solution,
\begin{equation}
j_{\pi_i} = \frac{8}{v_1^{(0)} v_2^{(0)} v_3^{(0)}} \qquad \forall \pi_i. \label{jb}
\end{equation}
This follows with \eqref{jb0} using that $\mathcal{V}_{\pi_i}^{(0)} = 3^{3/2} v_1^{(0)} v_2^{(0)} v_3^{(0)}$ and $\mathcal{V}^{(0)} = \frac{3^{3/2}}{8} v_1^{(0)2} v_2^{(0)2} v_3^{(0)2}$ on the covering torus, where we remind the reader that $j_{\pi_i} \equiv j_{\pi_i}^{(0)}$.
\\

\subsection{Solution}
\label{sec:ex-sol}

We are now ready to solve the equations \eqref{final-eom1}--\eqref{final-eom4} in the $T^6/{\mathbb{Z}_3^2}$ model. We first consider the components of the Einstein equation \eqref{final-eom4}. Recall that they are stated in real coordinates. Using \eqref{r}, \eqref{sm-metric} and Table \ref{tab-proj}, they become:
\begin{align}
R^{(1)}_{11} &= - \frac{1}{2} g^{(0)mn} \partial_1^2 g^{(1)}_{mn}+ g^{(0)mn} \partial_m \partial_1 g^{(1)}_{n1} -\frac{1}{2} \nabla^2 g_{11}^{(1)} \nll
 = \frac{4}{w^{(0)}} \partial_1^2 w^{(1)} + \frac{2}{\tau^{(0)}} \partial_1^2 \tau^{(1)} - \frac{v_1^{(0)2}}{2 \tau^{(0)}}\sum_{\substack{\pi_\pm,\pi_{31},\\\pi_{13},\pi_{12},\pi_{21}}}\!\!\!\! \left( j_{\pi_i}- \delta(\pi_i) \right)  + \frac{v_1^{(0)2}}{ \tau^{(0)}}\sum_{\substack{\pi_0,\pi_{23},\\\pi_{32}}}\!\! \left( j_{\pi_i}- \delta(\pi_i) \right), \label{einstein11} \\
R^{(1)}_{22} &= - \frac{1}{2} g^{(0)mn} \partial_2^2 g^{(1)}_{mn}+ g^{(0)mn}\partial_m \partial_2 g^{(1)}_{n2}-\frac{1}{2} \nabla^2 g_{22}^{(1)} \nll
 =  \frac{4}{w^{(0)}} \partial_2^2 w^{(1)} + \frac{2}{\tau^{(0)}} \partial_2^2 \tau^{(1)} + \frac{v_1^{(0)2}}{2 \tau^{(0)}}\sum_{\substack{\pi_\pm,\pi_{31},\\\pi_{13},\pi_{12},\pi_{21}}}\!\!\!\! \left( j_{\pi_i}- \delta(\pi_i) \right)  - \frac{v_1^{(0)2}}{ \tau^{(0)}}\sum_{\substack{\pi_0,\pi_{23},\\\pi_{32}}}\!\! \left( j_{\pi_i}- \delta(\pi_i) \right), \label{einstein22} \\
R^{(1)}_{12} &= - \frac{1}{2} g^{(0)mn} \partial_1\partial_2 g^{(1)}_{mn}+\frac{1}{2} g^{(0)mn} \left( \partial_m \partial_1 g^{(1)}_{n2}+\partial_m \partial_2 g^{(1)}_{n1}\right) -\frac{1}{2} \nabla^2 g_{12}^{(1)} \nll = \frac{4}{w^{(0)}} \partial_1\partial_2 w^{(1)} + \frac{2}{\tau^{(0)}} \partial_1\partial_2 \tau^{(1)} + 
\frac{\sqrt{3}v_1^{(0)2}}{2 \tau^{(0)}} \Big( \sum_{\substack{\pi_+,\pi_{13},\\\pi_{12}}}\!\!\!\! \left( j_{\pi_i}- \delta(\pi_i) \right)  - \!\!\sum_{\substack{\pi_-,\pi_{31},\\\pi_{21}}}\!\!\!\! \left( j_{\pi_i}- \delta(\pi_i) \right) \Big), \label{einstein12} \\
R^{(1)}_{13} &= - \frac{1}{2} g^{(0)mn} \partial_1\partial_3 g^{(1)}_{mn}+\frac{1}{2} g^{(0)mn} \left( \partial_m \partial_1 g^{(1)}_{n3}+\partial_m \partial_3 g^{(1)}_{n1}\right) -\frac{1}{2} \nabla^2 g_{13}^{(1)} \nll = \frac{4}{w^{(0)}} \partial_1\partial_3 w^{(1)} + \frac{2}{\tau^{(0)}} \partial_1\partial_3 \tau^{(1)}. \label{einstein13}
\end{align}
Here, we have only written down the first few components of the Einstein equation because the remaining equations follow from the above by simple replacements: The 14, 23 and 24 components are obtained from \eqref{einstein13} by replacing the indices $\left\{1,3\right\}$ by $\left\{1,4\right\}$, $\left\{2,3\right\}$ or $\left\{2,4\right\}$. All other components of the Einstein equation then follow from the above equations by ``exchanging'' the three 2-tori, i.e., by shifting the indices such that $m,n \to m+2,n+2$ (in $g_{mn}^{(1)}$ and $\partial_m$) and $a,b\to a+1,b+1$ (in $v_a^{(0)}$ and $\pi_{ab}$).
For example, the equation for $R^{(1)}_{44}$ follows from \eqref{einstein22} by replacing $\partial_2, g_{22}^{(1)},g_{n2}^{(1)} \to \partial_4, g_{44}^{(1)},g_{n4}^{(1)}$ and $v_1^{(0)}\to v_2^{(0)}$, $\pi_{ab} \to \pi_{a+1,b+1}$. Repeating the index shifts once again then yields the equation for  $R^{(1)}_{66}$, etc.

To simplify things, we will now solve this system under the temporary assumption that all image sources are absent, i.e., we only consider the O-plane wrapping $\pi_0$ in Table \ref{tab-pos}. We will later construct the full solution by symmetrizing the simplified solution with respect to the two $\mathbb{Z}_3$ symmetries $Q$ and $T$. According to Table \ref{tab-pos}, the O-plane wrapping $\pi_0$ is parallel to $y^2$, $y^4$, $y^6$. The fields sourced by it should therefore only depend on $y^1$, $y^3$, $y^5$, i.e., $\partial_2 w^{(1)}=\partial_2 \tau^{(1)} = \ldots = 0$. Since the configuration is invariant under exchanging the three 2-tori, we furthermore assume that
\begin{equation}
g^{(0)11} g^{(1)}_{11} = g^{(0)33} g^{(1)}_{33} = g^{(0)55} g^{(1)}_{55}, \qquad g^{(0)22} g^{(1)}_{22} = g^{(0)44} g^{(1)}_{44} = g^{(0)66} g^{(1)}_{66}
\end{equation}
and that $g_{mn}^{(1)}=0$ if $m\neq n$. As we will see momentarily, this ansatz is correct.

Under the above assumptions, \eqref{einstein12} is trivially satisfied. Using \eqref{jb}, the remaining Einstein equations simplify to
\begin{align}
0 &= - \frac{4}{w^{(0)}} \partial_1^2 w^{(1)} - \frac{1}{2} g^{(0)11} \partial_1^2 g^{(1)}_{11} - \frac{3}{2} g^{(0)22} \partial_1^2 g^{(1)}_{22} 
-\frac{1}{2} \nabla^2 g_{11}^{(1)} - \frac{2}{\tau^{(0)}} \partial_1^2 \tau^{(1)} \nl -\frac{v_1^{(0)}}{v_2^{(0)}v_3^{(0)}\tau^{(0)}}\, \rho, \label{simple-eom1} \\
0 &= -\frac{1}{2} \nabla^2 g_{22}^{(1)} + \frac{v_1^{(0)}}{v_2^{(0)}v_3^{(0)}\tau^{(0)}}\, \rho, \\
0 &= - \frac{4}{w^{(0)}} \partial_1\partial_3 w^{(1)} - \frac{3}{2} g^{(0)22} \partial_1\partial_3 g^{(1)}_{22} - \frac{1}{2} g^{(0)11} \partial_1\partial_3 g^{(1)}_{11} - \frac{2}{\tau^{(0)}} \partial_1\partial_3 \tau^{(1)}
\end{align}
with the source term
\begin{equation}
\rho = 8-\sum_{\substack{m_a\\\in\{1,2\}}} \delta\left(y^1-\tfrac{m_1}{2}\right)\delta\left(y^3-\tfrac{m_2}{2}\right)\delta\left(y^5-\tfrac{m_3}{2}\right). \label{rho}
\end{equation}
Recall now that we also need to solve the $F_2$ Bianchi identity, the dilaton equation and the warp factor equation, which are given by \eqref{final-eom1}--\eqref{final-eom3}. Under the above assumptions, they simplify to
\begin{align}
\d F_2^{(1)} &= 4 \rho\, \d y^1 \w \d y^3 \w \d y^5, \\
\nabla^2 \tau^{(1)} &= -\frac{3}{v_1^{(0)}v_2^{(0)}v_3^{(0)}}\, \rho, \\
\nabla^2 w^{(1)} &= \frac{w^{(0)}}{v_1^{(0)}v_2^{(0)}v_3^{(0)}\tau^{(0)}}\, \rho. \label{simple-eom2}
\end{align}
To solve these equations, we now introduce a function $\beta$ and set\footnote{Strictly speaking, $\tau^{(1)}$, $w^{(1)}$, etc.~are only determined by the function $\beta$ up to constants. The latter should be fixed by the equations of motion at next-to-leading order because there $\tau^{(1)}$, $w^{(1)}$, etc.~appear without derivatives. We refrain from doing so here because these constants are unimportant for the following discussion.}
\begin{equation}
\frac{g_{11}^{(1)}}{v_1^{(0)2}} = -\frac{g_{22}^{(1)}}{v_1^{(0)2}} = \frac{2}{3}\frac{\tau^{(1)}}{\tau^{(0)}} =  -2\frac{w^{(1)}}{w^{(0)}} = - \frac{2 \beta}{v_1^{(0)}v_2^{(0)}v_3^{(0)}\tau^{(0)}} \label{b0}
\end{equation}
and
\begin{equation}
F_2^{(1)}=-4 \star_6^{(0)}\! \left(\d \beta \w \d y^2 \w \d y^4 \w \d y^6\right). \label{f2}
\end{equation}
Note that the latter is consistent with \eqref{loc2b}.
Substituting the expressions into \eqref{simple-eom1}--\eqref{simple-eom2}, we observe that all equations reduce to a single Poisson equation:
\begin{equation}
\nabla^2 \beta = \rho. \label{beta1}
\end{equation}
Solving this  in flat space
would be a straightforward exercise.
However, in our case, the transverse space at each point on the O-plane worldvolume is a 3-torus, and so we have to take into account an infinite sum over image sources to ensure that $\beta$ is periodic with respect to the identifications \eqref{torus}.

For the familiar case of a 2-torus, the resulting Green function is given by the logarithm of a Jacobi theta function (see, e.g., \cite{Polchinski:1998rq, ooguri}). For the 3-torus, the Green function is less well known. One possibility is to write the solution in terms of a Fourier series. To this end, consider a unit torus parametrized by $\vec y= (y^1,y^3,y^5)^T\in [0,1]^3$ and a field $\phi$ satisfying $\nabla^2 \phi = 1 -\delta(\vec y)$. We can then write $1 -\delta(\vec y)= \sum_{\vec k\in\mathbb{Z}^3} \e^{2\pi i\vec k\vec y}\left(\delta_{\vec k0}-1 \right)= - \sum_{\vec k\in\mathbb{Z}^3\setminus\{0\}} \e^{2\pi i\vec k\vec y}$.
The formal solution is, up to a constant, $\phi= \sum_{\vec k\in\mathbb{Z}^3\setminus\{0\}} \e^{2\pi i\vec k\vec y}/(4\pi^2 k^2)$.
Using this, we find\footnote{Recall that the torus metric \eqref{sm-metric} contains $v_a^{(0)}$ factors. To obtain the second line, we used that $\sum_{k\in\mathbb{Z}\setminus\{0\}} \frac{1}{f(k)}\left[\exp(2\pi i k (y-1))+\exp(2\pi i k (y-\frac{1}{2}))\right]
= \sum_{k\in\mathbb{Z}\setminus\{0\}} \frac{2}{f(2k)}\exp(4\pi i k y) $.}
\begin{align}
\beta(\text{Re}(\vec z)) &= \sum_{\substack{m_a\\\in\{1,2\}}}\, \sum_{\vec k\in\mathbb{Z}^3\setminus\{0\}} \frac{\exp\left(2\pi i \vec k\cdot \left(\text{Re}(\vec z)-\frac{\vec m}{2}\right)\right)}{4\pi^2 k^2} +\text{const.} \nll = 2 \sum_{\vec k\in\mathbb{Z}^3\setminus\{0\}} \frac{\exp\left(4\pi i \vec k \cdot\text{Re}(\vec z)\right)}{4\pi^2 k^2} +\text{const.}.
\label{beta}
\end{align}
Here, we used the notation $\vec m = (m_1,m_2,m_3)^T$, $\vec k = (k_1,k_2,k_3)^T$ (with $k^2 = k_1^2 g^{11(0)}+k_2^2 g^{33(0)}+k_3^2 g^{55(0)} = k_1^2/v_1^{(0)2}+k_2^2/v_2^{(0)2}+k_3^2/v_3^{(0)2}$) and
$\vec z = (z_1,z_2,z_3)^T$ (with $y^1 = \text{Re}(z_1)$, $y^3 = \text{Re}(z_2)$ and $y^5 = \text{Re}(z_3)$, cf.~\eqref{coord}). One checks that the above expression is manifestly invariant under $z_a\to z_a +1$, $z_a\to z_a +\alpha$, in agreement with the periodic identifications \eqref{torus}.

However, it turns out that the formal Fourier series does not converge such that one has to regularize the Green function \cite{Shandera:2003gx} (see also \cite{Andriot:2019hay}). A regularized expression in terms of Jacobi theta functions was proposed in \cite{Shandera:2003gx} and passes some consistency checks \cite{Andriot:2019hay}. Using the result of \cite{Shandera:2003gx}, we can write
\begin{equation}
\beta(\text{Re}(z_a)) = - 2 \int_0^\infty \d s \left[ 1 - \prod_{a=1}^3 \theta_3\! \left( 2 \text{Re}(z_a), \exp\left({-\frac{4\pi^2s}{v_a^{(0)2}}}\right) \right) \right] +\text{const.},
\label{beta2}
\end{equation}
where $\theta_3$ is defined as
\begin{equation}
\theta_3(b;\e^{-a}) = \sum_{n=-\infty}^{\infty} \e^{-an^2+2\pi ibn}.
\end{equation}

We also note for later convenience that, at distances $r\to 0$ very close to the O-plane, the behavior of $\beta$ approaches the non-compact solution (see \cite{Andriot:2019hay} for an explicit check):
\begin{align}
\beta(r\to 0) = \frac{v_1^{(0)}v_2^{(0)}v_3^{(0)}}{4\pi r} + \text{const.}. \label{betar}
\end{align}
Here, $r$ is the distance measured with $g_{mn}^{(0)}$. For example, for a source localized at $y^1=y^3=y^5=0$, we would have $r=(g_{11}^{(0)}(y^1)^2+g_{33}^{(0)}(y^3)^2+g_{55}^{(0)}(y^5)^2)^{1/2} = (v_{1}^{(0)2}(y^1)^2+v_{2}^{(0)2}(y^3)^2+v_{3}^{(0)2}(y^5)^2)^{1/2}$.

The final step is now to symmetrize our result with respect to $Q$ and $T$ in order to include the backreaction of the image O-planes ($\pi_i\in\pi_\pm,\pi_{ab}$) that we neglected so far.
To this end, recall the internal metric in complex coordinates defined in \eqref{complex}. In this notation, our simplified result \eqref{b0} amounts to (cf.~\eqref{real1}--\eqref{uvexp})
\begin{align}
v_a &= v_a^{(0)} n^{1/4} + \mathcal{O}(n^{-7/4}), \\ u_a &= \frac{u_a^{(1)}}{n^{1/2}} + \mathcal{O}(n^{-3/2}) = - \frac{v_a^{(0)2}}{n^{1/2}v_1^{(0)}v_2^{(0)}v_3^{(0)}\tau^{(0)}} \,\beta\left(\text{Re}(z_b)\right)+ \mathcal{O}(n^{-3/2}).
\end{align}
Since $u_a$ is required to behave under actions of $Q$ and $T$ as in \eqref{ua}, we can guess the properly symmetrized version:
\begin{align}
u_a^{(1)} &= - \frac{v_a^{(0)2}}{v_1^{(0)}v_2^{(0)}v_3^{(0)}\tau^{(0)}}\sum_{q,t=0,1,2} \alpha^{4aq+4t} \beta\left(\text{Re}\left(Q^q T^t z_b\right)\right) \nll = - \frac{v_a^{(0)2}}{v_1^{(0)}v_2^{(0)}v_3^{(0)}\tau^{(0)}}\sum_{q,t=0,1,2} \alpha^{4aq+4t} \beta\left(\text{Re}\left(\alpha^{2bq+2t} z_b\right)\right). \label{ua1}
\end{align}
Here, by $\beta\left(\text{Re}\left(Q^q T^t z_b\right)\right)$ we mean that the function $\beta$ defined in \eqref{beta2} should be evaluated with argument $\text{Re}\left(Q^q T^t z_b\right)$ instead of $\text{Re}(z_b)$, where $Q$ and $T$ act on $z_b$ as in \eqref{qt}. For the second line, we used that $\beta$ is invariant under shifts of its argument by $\frac{1}{2}$.
Also note the factor $\alpha^{4aq+4t}$ in the sum, which ensures that $u_a \sim \alpha^{-4}$ under actions of $T$ and $u_a \sim \alpha^{-4a}$ under actions of $Q$ (as required by \eqref{ua}).

\begin{figure}[t!]
\centering
\includegraphics[trim = 0mm 0mm 0mm 0mm, clip, width=0.4\textwidth]{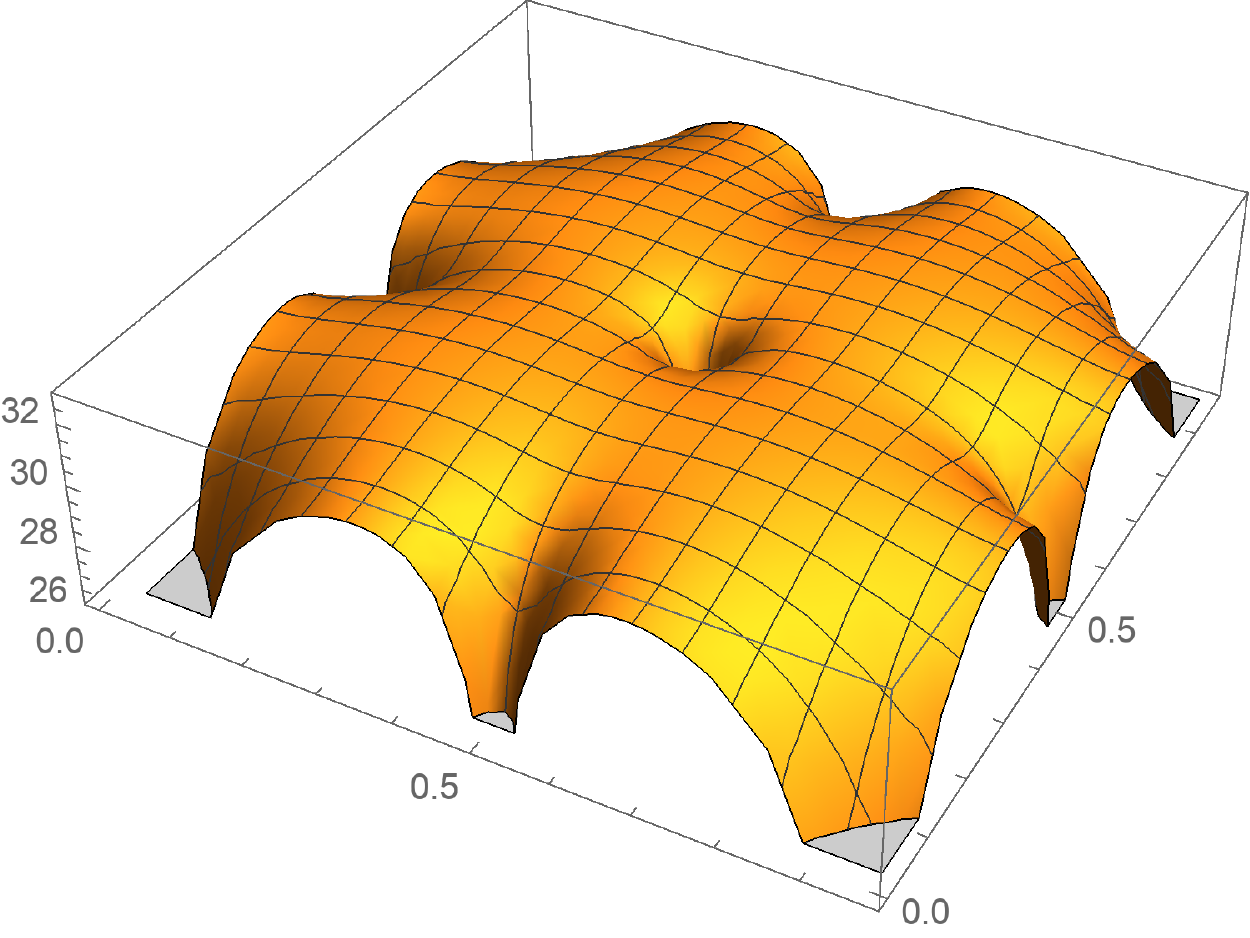}\qquad\qquad\includegraphics[trim = 0mm 0mm 0mm 0mm, clip, width=0.4\textwidth]{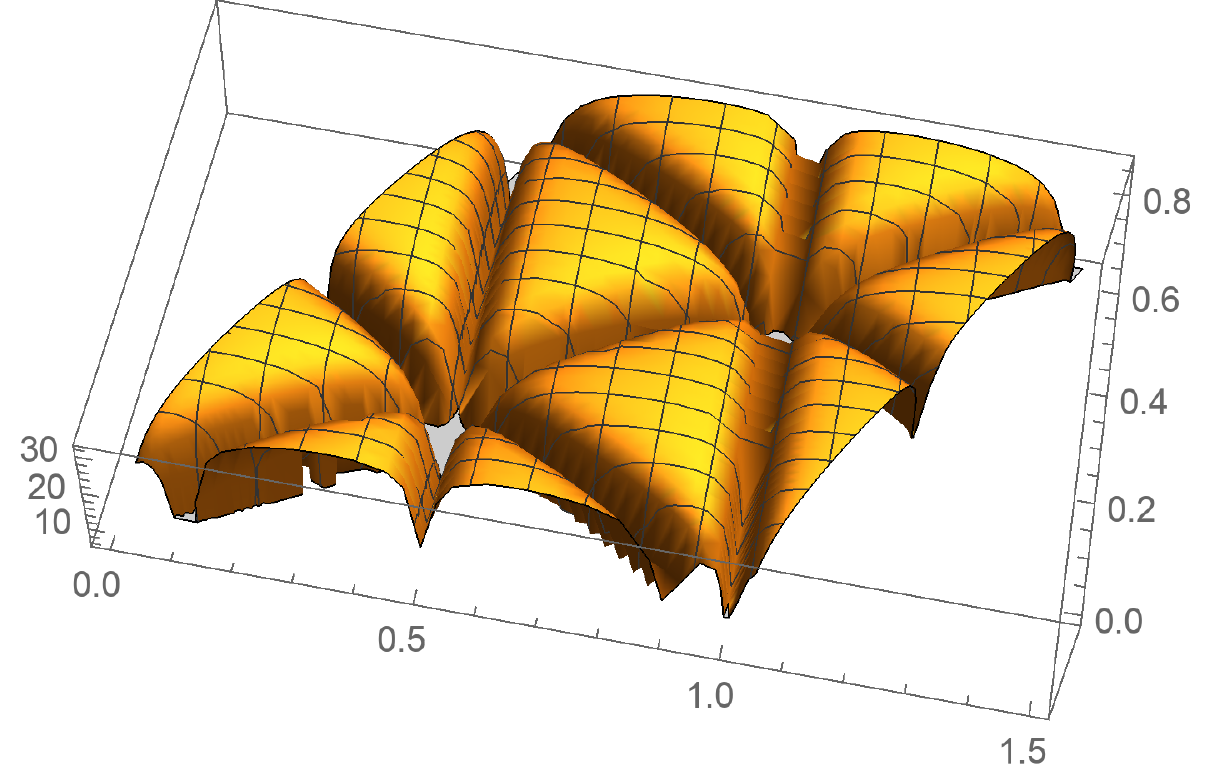}
\put(-417,70){$\scriptstyle{\tau}$}
\put(-358,10){$\scriptstyle{y^1}$}
\put(-242,40){$\scriptstyle{y^3}$}
\put(-186,45){$\scriptstyle{\tau}$}
\put(-105,8){$\scriptstyle{y^1}$}
\put(-4,50){$\scriptstyle{y^2}$}

\caption{
Solution for $\tau$ with $n=100$ on the 2-tori parametrized by $y^1$, $y^3$ (left) and by $y^1$, $y^2$ (right) with the remaining $y^m=0$. The $\mathcal{O}(1)$ expansion coefficients $\tau^{(0)}$, $v_a^{(0)}$ are set to 1 for concreteness.
\label{tau1}}
\end{figure}
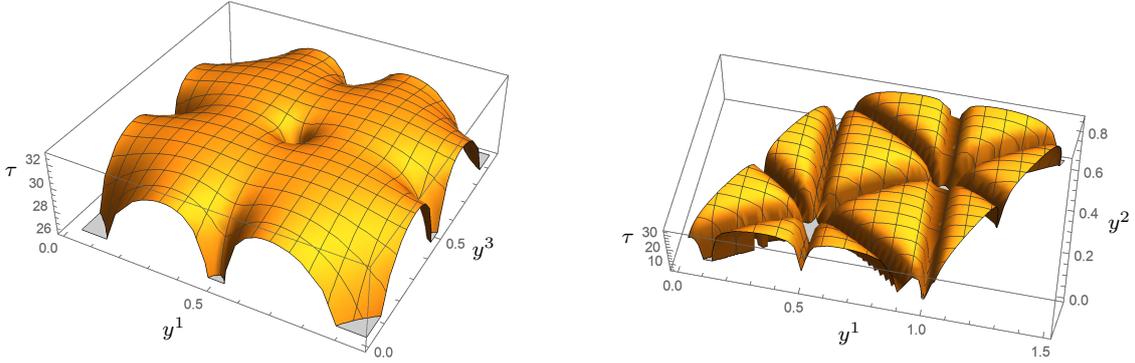

Similarly, $w$ and $\tau$ need to be invariant under $Q$ and $T$, and so we have
\begin{equation}
\frac{w^{(1)}}{w^{(0)}} = -\frac{\tau^{(1)}}{3\tau^{(0)}} = \frac{1}{v_1^{(0)}v_2^{(0)}v_3^{(0)}\tau^{(0)}} \sum_{q,t=0,1,2} \beta\left(\text{Re}\left(\alpha^{2aq+2t} z_a\right)\right). \label{sol0}
\end{equation}
The resulting behavior of the dilaton on the covering torus is plotted in Fig.~\ref{tau1}. Note in particular how the backreaction reflects the presence of the various O-plane images (cf.~Fig.~\ref{fund}). Because of \eqref{sol0}, the warp factor behaves in the same way (up to a constant and a proportionality factor).

Finally, the same argument also fixes $F_2$:
\begin{align}
F_2^{(1)} = -\frac{i}{2} \sum_{q,t=0,1,2} \star_6^{(0)}& \bigg[\d \beta\left(\text{Re}\left(\alpha^{2aq+2t} z_a \right)\right) \w \d ( \alpha^{2q+2t} z_1- \alpha^{-2q-2t}z_1^*) \w \nl\!\!\!\! \d (\alpha^{4q+2t}z_2-\alpha^{-4q-2t}z_2^*) \w \d (\alpha^{2t}z_3-\alpha^{-2t}z_3^*)\bigg].
\end{align}
Note that $v_a$ is a constant and therefore already invariant under $Q$ and $T$ such that we need not modify it further. As a double-check, we can now substitute the above solution for $F_2^{(1)}$, $w^{(1)}$, $\tau^{(1)}$ and $g_{mn}^{(1)}$ into \eqref{final-eom1}--\eqref{final-eom4} (where we have to express the $z_a$ in terms of the real coordinates $y^m$ using \eqref{coord}). One then verifies that, taking the source distribution as in Tables \ref{tab-pos}, \ref{tab-proj}, all equations of motion are indeed satisfied as expected.
\\

\section{Scalar Potential}
\label{sec:scalar}

Now that we established the existence of a solution including the O-plane backreaction, an important question is whether its low-energy physics differs from the one of the smeared solution.
In particular, we would like to understand whether the backreaction corrections significantly affect the 4d scalar potential, which was derived in the smeared approximation in \cite{Grimm:2004ua}.
We will argue in this section that such corrections are actually suppressed by $1/\sqrt{n}$. Hence, remarkably, the low-energy EFT obtained in the smeared approximation becomes exact in the large-$n$ (i.e., large-volume, small-$g_s$) limit.
Aside from backreaction effects, we will also discuss string ($\alpha^\prime$ and $g_s$) corrections to the scalar potential. We will argue that these are suppressed in the large-$n$ limit as well.
Note that the following discussion applies generally to the AdS solutions studied in this paper, although we will use the orbifold example of the previous section to illustrate some crucial points.

\subsection{Leading-order Potential}
\label{sec:scalar1}

We derive the scalar potential by performing a dimensional reduction of the 10d type IIA supergravity action.
This yields
\begin{equation}
S_\text{IIA} \supset 2\pi \int \d^4 x \sqrt{-g_4} \left(\mathcal{V}_\text{w} R_{\mu\nu}g^{\mu\nu} - V\right), \qquad \mathcal{V}_\text{w} = \int \d^6 y \sqrt{g_6} \, w^2\tau^2. \label{wv}
\end{equation}
Here, $g_4 =\det(g_{\mu\nu})$, $g_6 =\det(g_{mn})$, $\mathcal{V}_\text{w}$ is the warped volume\footnote{Note that $\mathcal{V}_\text{w}$ differs from the string-frame volume $\mathcal{V} = \int \d^6 y \sqrt{g_6}$ by a factor $w^2\tau^2$.} and
\begin{align}
V = \int \d^6 y \sqrt{g_6} \, w^4 &\bigg( 12 \frac{\tau^2}{w^2} (\partial w)^2 + 8\frac{\tau^2}{w}\nabla^2 w - \tau^2 R_{mn}g^{mn} - 4 (\partial \tau)^2 + \frac{1}{2}\tau^2 |H_3|^2 \nl + \frac{1}{2} \sum_{q=0}^6 |F_q|^2 - 2\sum_i \tau \delta(\pi_i) \bigg). \label{pot}
\end{align}
Alternatively, the same expression can be obtained by integrating the trace of the energy-momentum tensor $T_\mu^\mu$ over the internal space.
Note that the last term on the right-hand side is due to the DBI action of the O6-planes, while the other terms come from the 10d bulk action. Also note that the derivatives in \eqref{pot} are with respect to the internal coordinates $y^m$ only. Here and in the following, we ignore terms involving external derivatives (such as $\partial_\mu \tau$, $H_{3,\mu mn}$, etc.). These would yield the kinetic terms for the various 4d fields but are not relevant for the scalar potential, which is the focus of this section.

The localized O6-plane term in \eqref{pot} can be eliminated in terms of a bulk term using the Bianchi identity \eqref{f2bianchi} and partial integration:
\begin{align}
V = \int \d^6 y \sqrt{g_6} \, w^4 &\bigg( 12 \frac{\tau^2}{w^2} (\partial w)^2 + 8\frac{\tau^2}{w}\nabla^2 w - \tau^2 R_{mn}g^{mn} - 4 (\partial \tau)^2 + \frac{1}{2}\tau^2 |H_3|^2 \nl + \frac{1}{2} \sum_{q=0}^6 |F_q|^2 \bigg) - \sum_i \int \text{vol}_{\pi_i} \w  \left[\d (w^4 \tau) \w F_2 + w^4 \tau H_3F_0\right]. \label{pot2}
\end{align}
This form of $V$ is equivalent to \eqref{pot} at the supergravity level. It is required, however, to match the result of the dimensional reduction with the corresponding $F$-term potential in the 4d $\mathcal{N}=1$ supergravity formulation \cite{DeWolfe:2005uu}. Writing $V$ this way will also be more convenient for the analysis of corrections in the next section.

We now go to 4d Einstein frame by redefining $g_{\mu\nu} = g_{\mu\nu,E} (2\pi\mathcal{V}_\text{w})^{-1}$ in \eqref{wv}.
This yields the 4d effective action in Planck units:
\begin{equation}
S_{4d} = \int \d^4 x \sqrt{-g_E} \left(R_E - V_E \right)
\end{equation}
with 4d scalar potential $V_E = V/(2\pi\mathcal{V}_\text{w}^2)$.

The final step is to substitute our large-$n$ expansion \eqref{ansatz0}--\eqref{ansatz} into the scalar potential $V_E$.\footnote{This is valid off-shell as long as we do not impose the equations of motion. For this reason, we have to reintroduce the zeroth-order piece $F_6^{(0)}$ in $F_6$, which we set to zero earlier on-shell.} As mentioned before, this expansion is valid everywhere on the compact space except very close to the O-plane sources. Strictly speaking, we are therefore not allowed to expand \eqref{pot2}, as it involves an integration over the full 6d space. However, let us ignore this for the moment and assume that \eqref{ansatz0}--\eqref{ansatz} holds everywhere. We will argue in the next section that the error $\delta V_E$ created by this assumption becomes negligible in the large-$n$ limit, $\delta V_E/V_E \to 0$.

We thus obtain
\begin{align}
V_E &= \frac{w^{(0)2}}{2\pi n^{9/2} \mathcal{V}_\text{w}^{(0)} \tau^{(0)2}} \bigg( \frac{1}{2}\tau^{(0)2} |H_3^{(0)}|^2 + \frac{1}{2} \sum_{q=0}^6 |F_q^{(0)}|^2
\bigg) \nl - \frac{1}{2\pi n^{9/2} \mathcal{V}_\text{w}^{(0)2}} \sum_i \int \text{vol}^{(0)}_{\pi_i} \w w^{(0)4} \tau^{(0)} H^{(0)}_3F_0 + \mathcal{O}(n^{-5}) \label{lopotential}
\end{align}
with
\begin{equation}
\mathcal{V}_\text{w}^{(0)} = \int \d^6 y \sqrt{g_6^{(0)}} \, w^{(0)2}\tau^{(0)2}, \label{v0}
\end{equation}
where we used \eqref{r} and that total derivatives integrate to zero.

Crucially, this result is equivalent to what we would have obtained from a dimensional reduction of the smeared setup. Indeed, substituting the smeared ansatz \eqref{smeared0} into \eqref{pot2} and using \eqref{v0}, we would again arrive at \eqref{lopotential}.
We thus observe that the 4d scalar potential agrees with the corresponding smeared expression at leading order in the large-$n$ expansion. In order to make explicit that \eqref{lopotential} matches the usual $F$-term scalar potential, one would have to
expand all fields in terms of the K\"{a}hler and complex-structure moduli. We refrain from reviewing this procedure here and refer to \cite{Grimm:2004ua, DeWolfe:2005uu} for detailed discussions. As stated before, we also omitted an explicit derivation of the kinetic terms for the various moduli. According to the expansion \eqref{ansatz0}--\eqref{ansatz}, we again expect that they will agree with the corresponding smeared expressions up to subleading $1/n$ corrections.

Our calculation in this section confirms the common lore that backreaction becomes negligible in the large-volume, small-$g_s$ limit. We conclude that the interesting properties of the smeared DGKT solutions and their non-supersymmetric cousins (i.e., tree-level moduli stabilization and AdS/KK scale separation) carry over to their backreacted counterparts. Indeed, due to the identical scalar potential at large $n$, we expect that the moduli are stabilized at the same vevs as in the smeared case, up to subleading corrections.
Similarly, we expect that the AdS and KK scales are only corrected by subleading terms at large $n$: The AdS scale is related to the on-shell potential $\langle V_E\rangle$ by the Einstein equations and therefore identical to its smeared value at leading order. The KK scale is related to the moduli vevs controlling the cycle volumes, which are at leading order again determined by the smeared potential.
\\

\subsection{Corrections}
\label{sec:scalar2}

We now estimate the various corrections to the scalar potential $V_E$. This is complicated by the fact that
the large-$n$ expansion of the 10d fields breaks down close to the O-planes.
We will therefore separately estimate corrections to $V_E$ arising in the near-source regions and corrections arising in the ``bulk'', i.e., sufficiently far away from the O-planes where our solution is valid.

We first consider the bulk corrections. Since our large-$n$ expansion is trustworthy there, we can estimate such corrections simply by counting powers of $n$.
As we have seen above, the leading potential \eqref{lopotential} scales like $n^{-9/2}$.
According to \eqref{ansatz0}--\eqref{ansatz}, the backreaction corrections are at least suppressed by a factor $n^{-1/2}$ and therefore appear earliest at the order $n^{-5}$. Furthermore, there are perturbative $\alpha^\prime$ and $g_s$ corrections. One can verify that $m$-derivative terms are at least suppressed by a factor $n^{-(m-6)/4}$ and that loop corrections come with extra powers of $\tau^{-2} \sim n^{-3/2}$ compared to the leading terms.\footnote{Note that the suppression can be much stronger for some higher-derivative terms. For example, $R^4 \sim n^{-2}$ (with $R^4$ any scalar combination of 4 Riemann tensors) and $(|H_3|^2)^4\sim n^{-6}$ are both 8-derivative terms but the latter is suppressed by a factor $n^{-9/2} \ll n^{-(m-6)/4}$ compared to leading terms like $|H_3|^2\sim n^{-3/2}$.} We conclude that $l$-loop, $m$-derivative corrections to the leading potential scale like
\begin{equation}
\lesssim n^{-9/2-3l/2-(m-6)/4}.
\end{equation}
Hence, as already noted in \cite{DeWolfe:2005uu}, perturbative string corrections in the bulk are parametrically suppressed at large $n$. For example, assuming that the earliest $\alpha^\prime$ corrections appear at the 8-derivative and zero-loop level, they would be of the order $n^{-5}$ or smaller.

\begin{figure}[t!]
\centering
\!\!\!\includegraphics[trim = 0mm 0mm 0mm 0mm, clip, width=0.48\textwidth]{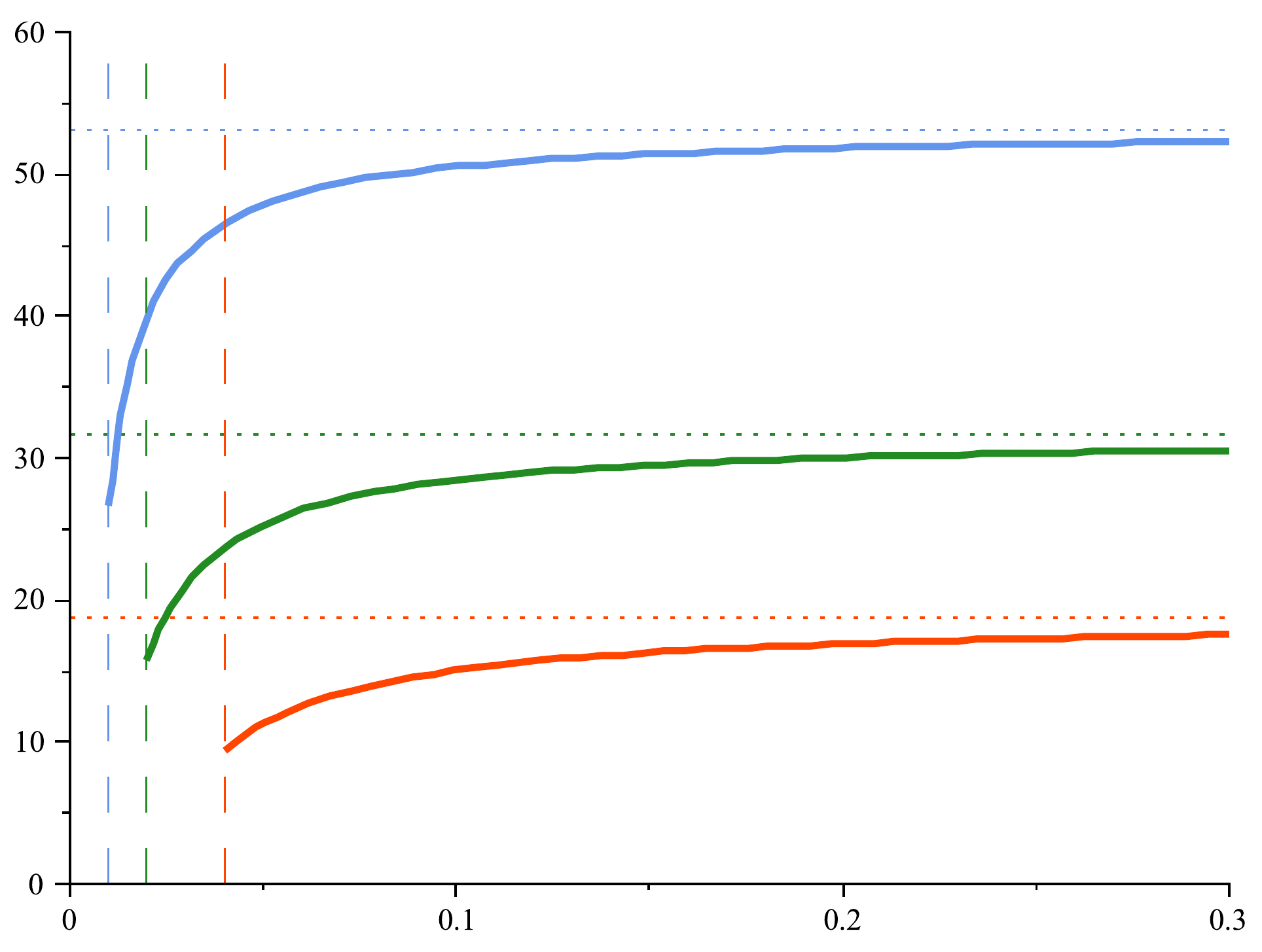}\quad\,\,\includegraphics[trim = 0mm 0mm 0mm 0mm, clip, width=0.48\textwidth]{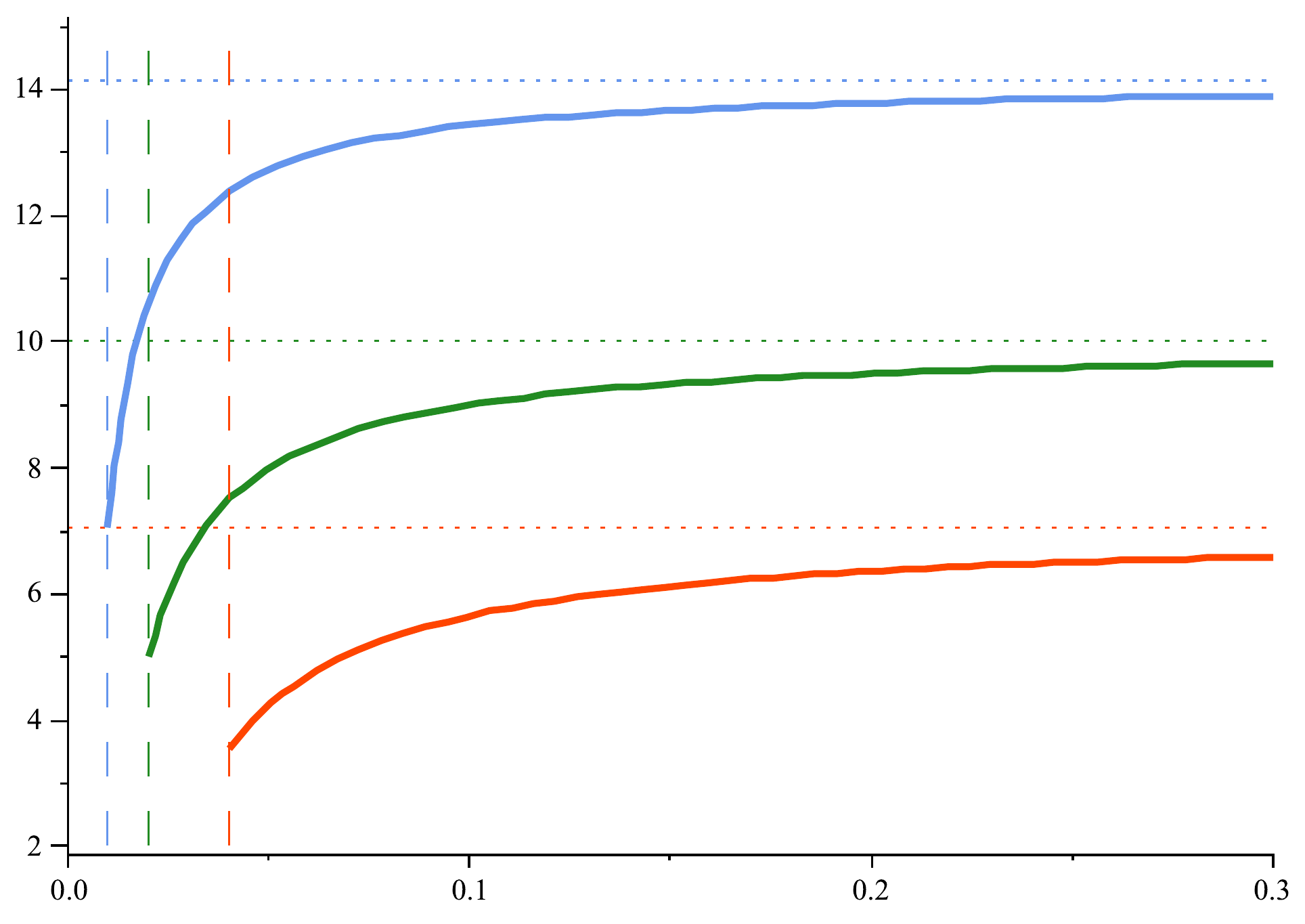}
\put(-456,148){$\scriptstyle{\tau}$}
\put(-228,148){$\scriptstyle{g_{11}}$}
\put(-18,-5){$\scriptstyle{r}$}
\put(-250,-5){$\scriptstyle{r}$}

\caption{Near-source behavior of $\tau$ and $g_{11}$
for $n=50$ (orange), $n=100$ (green) and $n=200$ (blue), in string units. We ignore $\mathcal{O}(1)$ expansion coefficients and the effect of further sources at $r\neq 0$ as in \eqref{s1}, \eqref{s2}. Dotted lines indicate the corresponding smeared solution.
The NLO terms become significant in the regions $r\sim n^{-1}$ left of the dashed lines. For concreteness, we took the boundaries to be the values of $r$ at which the NLO terms grow to half of the size of the leading term.
\label{source}}
\end{figure}

We now turn to the unreliable regions near the O-planes. To see that our solution is not valid there, consider the neighborhood of an O-plane localized at $y^1=y^3=y^5=0$ in the simple model studied in Section \ref{sec:example}. According to \eqref{ansatzw}, \eqref{ansatz}, \eqref{b0} and \eqref{betar}, the dilaton and the transverse metric in the near-source region are then given by
\begin{align}
\tau &= n^{3/4} - \frac{1}{n^{1/4} r} + \mathcal{O}(n^{-5/4}), \label{s1} \\
g_{11} &= g_{33} = g_{55} = n^{1/2} - \frac{1}{n^{1/2} r} + \mathcal{O}(n^{-3/2}), \label{s2}
\end{align}
and similar expressions can be derived for the other fields. Here, we ignored all $\mathcal{O}(1)$ factors for simplicity and $r=(g^{(0)}_{11}(y^1)^2+g^{(0)}_{33}(y^3)^2+g^{(0)}_{55}(y^5)^2)^{1/2}$ is the distance
to the O-plane as measured with $g_{mn}^{(0)}$.\footnote{There are several inequivalent notions of distance/length one may consider here, which may be confusing. The distance $l_{X}$ on the backreacted orientifold $X$ is given by extremizing $ \int \d s\, (g_{mn} \frac{\d y^m}{\d s} \frac{\d y^n}{\d s})^{1/2}$. Another possible measure is the distance $l_{X_0} = (n^{1/2}g_{mn}^{(0)} y^m y^n)^{1/2}$ on the un-backreacted toroidal orientifold $X_0$. Analogously, one may define the corresponding distances $l_X^E$ and $l_{X_0}^E$ with respect to the (un-)backreacted Einstein-frame metrics $\sqrt{\tau}g_{mn}$ and $n^{7/8}\sqrt{\tau^{(0)}}g^{(0)}_{mn}$. Finally, since the equations in Section \ref{sec:ex-sol} are expressed in terms of the metric $g_{mn}^{(0)}$, their solution naturally involves the distance $r = (g_{mn}^{(0)}y^my^n)^{1/2}$ on the corresponding torus with that metric
(cf.~\eqref{betar}). Sufficiently far away from an O-plane, the backreaction is negligible such that $l_{X} \approx l_{X_0} = n^{1/4}r$ and $l_{X}^E \approx l_{X_0}^E \sim n^{7/16}r$. Here, we are interested in the region very close to the sources, where we do not know the full geometry.
However, since the transverse metric and the dilaton scale like $g_{mn} \sim n^{1/2}$ and $\tau \sim n^{3/4}$ in the bulk and become smaller near the O-plane, we expect that, parametrically, $r \lesssim l_{X} \lesssim n^{1/4} r$ and $r \lesssim l_{X}^E \lesssim n^{7/16} r$.
\label{fn1}}
We now observe that, for $r \lesssim n^{-1}$ (in string units), the formally subleading terms in $\tau$ and $g_{mn}$ have the same order of magnitude as the leading ones such that our expansion breaks down, see Fig.~\ref{source}. In this region, our solution has to be glued to a local solution including the full non-linear backreaction.
Furthermore, the $\alpha^\prime$ and string-loop expansions appear to break down as well since $\tau \to 0$, $g_{mn} \to 0$ at $r\sim n^{-1}$, which implies that the string coupling and the curvature/field strengths diverge.\footnote{It is possible that the apparent breakdown of
the $\alpha^\prime$ and string-loop expansions is just an artifact of not taking into account the non-linear backreaction in this region. For example, as discussed at the end of Section \ref{sec:sol}, there may be a regular solution similar to the one in \cite{Saracco:2012wc} already at the level of the supergravity equations. 
On the other hand, finite-$r$ singularities are known to appear in many O-plane solutions in supergravity, even including the non-linear backreaction.
Indeed, the above expressions for $\tau$ and $g_{mn}$ suggest that the non-linear solution in our case would locally resemble an O6-plane in flat space. It is therefore plausible that string corrections will be locally relevant even in the non-linear solution.}

The above seems to indicate
that our solution can be trusted for $r\gtrsim n^{-1}$. However, it is plausible that it already breaks down at larger distances,
as we will now explain.
We display an example for a leading term and one for a subleading term in the scalar potential, again keeping only the dependence on $n$ and $r$:
\begin{equation}
|H_3|^2 \sim n^{-3/2} + \mathcal{O}(n^{-2}), \qquad \frac{(\partial \tau)^2}{\tau^2} \sim \frac{1}{n^{5/2}r^4} + \mathcal{O}(n^{-7/2}),
\end{equation}
where we used \eqref{ansatzh}, \eqref{ansatz} and \eqref{s1}.
Evidently, the second term is suppressed by a factor $1/(nr^4)$ compared to the first one. It is straightforward to verify that other subleading terms such as $(\partial w)^2/w^2$, etc.~follow the same rule.
This suggests that the large-$n$ expansion already breaks down at $r \sim n^{-1/4}$ rather than $r \sim n^{-1}$. 
Furthermore, $\alpha^\prime$ corrections involving higher derivatives (such as, e.g., $(\nabla^2 (\partial\tau)^2)/\tau^2$) are only suppressed by powers of $1/(\sqrt{n}r^2)$ as well (compared to classical terms like $(\partial \tau)^2/\tau^2$): Each derivative produces a factor $1/r$ when acting on a field and each pair of derivatives comes with an inverse metric $\sim n^{-1/2}$.
We do not know whether higher-derivative terms of the above type appear in the effective action. However, if they do,
this means that not only backreaction corrections but also $\alpha^\prime$ corrections become relevant at $r \sim n^{-1/4}$.
Note that this can happen even though all field strengths and curvatures remain small in string units ($\sim n^{-3/2}$) down to much smaller values $r\sim n^{-1}$.\footnote{This is not a contradiction, as a small field strength only ensures that higher powers but not necessarily derivatives of it will be subleading. In other words, unless there are specific cancellations, the validity of the $\alpha^\prime$ expansion requires field strengths to be small \emph{and} slowly varying, where the latter is violated for $r\lesssim n^{-1/4}$.}

It is therefore not clear whether $r\sim n^{-1}$ or $r\sim n^{-1/4}$ should be considered the distance at which our expansion breaks down. To be on the safe side, we will work with the more conservative assumption $r\sim n^{-1/4}$ in the following.

Let us now estimate whether the unreliable regions affect the validity of the smeared approximation in 10d and in 4d.
We expect that a 10d observer at energies below $M_s$ can resolve Einstein-frame distances $l_X^E \gtrsim 1$ in string units.
We can further estimate that the corresponding distances measured with $g^{(0)}_{mn}$ are $r \gtrsim r_\text{exp}$ with $n^{-7/16} \lesssim r_\text{exp} \lesssim n^{0}$ parametrically (see footnote \ref{fn1}).
On the other hand, backreaction/string effects are localized in the region $r \lesssim r_0$ with $n^{-1} \lesssim r_0 \lesssim n^{-1/4}$.
If $r_0 \lesssim r_\text{exp}$, the regions of large backreaction cannot be resolved at sub-stringy energies where 10d supergravity is a valid description. We then expect that a 10d observer finds the smeared solution to be a good approximation to the exact solution \emph{locally}, i.e., everywhere on the internal space. On the other hand, if $r_0 \gtrsim r_\text{exp}$, the regions of large backreaction can be probed. Consequently, the 10d solution is not everywhere approximated by the smeared one, even at large $n$. Since we do not know the full solution near the O-planes, we do not know which of the two possibilities applies in our case. However, in any case, one may still ask whether the smeared solution is a good approximation in an integrated sense, i.e., whether it approximates the 4d low-energy physics at energies below $M_\text{KK}$. We will now argue that this is indeed the case.

\begin{figure}[t!]
\centering

\includegraphics[trim = 0mm 50mm 0mm 60mm, clip, width=\textwidth]{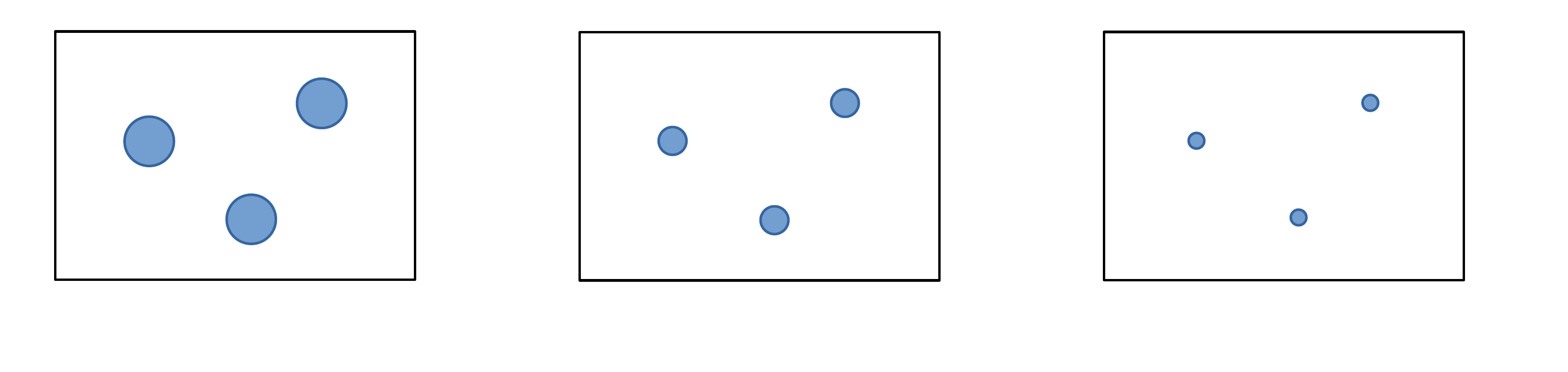}
\put(-403,5){$n=10^2$}
\put(-252,5){$n=10^3$}
\put(-100,5){$n=10^4$}

\caption{Schematic depiction of a 2d slice of the internal space (white rectangles) and regions $r\sim n^{-1/4}$ where our solution breaks down (blue shaded areas) for different $n$. In the unreliable regions, the solution has to be completed into a (possibly stringy) solution with resolved O-plane singularities.
\label{regions}}
\end{figure}

As explained above, the unreliable regions are
``tubes'' around the O6-planes with diameters $r_0 \lesssim n^{-1/4}$, see Fig.~\ref{regions}.
Nevertheless,
we pretended in Section \ref{sec:scalar1} that the large-$n$ expansion is valid there.
We now estimate the error created by this in our computation of the warped volume $\mathcal{V}_\text{w}$ (as defined in \eqref{wv}). 
The warped volume of the tube region evaluated on the smeared solution is
\begin{align}
\left.\int_{r \lesssim r_0} \d^6 y \sqrt{g_6}\, w^2\tau^2\right|_{\substack{\tau,w \sim n^{3/4}\\g_{mn}\sim n^{1/2}}}
 \sim n^{9/2} \int_{r \lesssim r_0} \d^6 y \sqrt{g_6^{(0)}} \sim n^{15/4}.
\end{align}
Let us again be conservative and assume that the
corrections to the smeared solution are a leading-order effect for all $r\lesssim n^{-1/4}$.
The error should then be of the order of the smeared tube volume itself:
\begin{align}
\delta \mathcal{V}_\text{w} & \lesssim n^{15/4}.
\end{align}
Crucially, this is a subleading correction to $\mathcal{V}_\text{w}$, which scales like $n^{9/2}$ at leading order. Hence, even though backreaction/string effects are large locally near the O-planes, we expect their contribution to $\mathcal{V}_\text{w}$ to be negligible because $\delta \mathcal{V}_\text{w}/\mathcal{V}_\text{w} \to 0$ in the large-$n$ limit.
It is therefore justified to approximate $\mathcal{V}_\text{w} = \mathcal{V}_\text{w}^{(0)} n^{9/2}+ \mathcal{O}(n^{15/4})$ as we did in the previous section.

An analogous argument can be made for the error $\delta V_E$ in our computation of the scalar potential. If our result \eqref{lopotential} is to make sense, we need to make sure that there are no large corrections to $V_E$ due to our ignorance about the tube regions. Recall that $V_E$ is given by integrating the (trace of the) energy-momentum tensor over the compact space, times a factor $\mathcal{V}_\text{w}^{-2}$ because of the usual conversion to 4d Einstein frame. Let us again assume that the error is of the order of the smeared energy-momentum in the tube regions.
We thus find
\begin{equation}
\delta V_E \sim \frac{\delta V}{\mathcal{V}_\text{w}^2} -2V\frac{ \delta \mathcal{V}_\text{w}}{\mathcal{V}_\text{w}^3} \lesssim n^{-21/4}.
\end{equation}
This is a subleading correction to the potential \eqref{lopotential}, which scales like $n^{-9/2}$. We thus see that, although corrections to our solution are locally relevant near the O-planes, the \emph{integrated} effect of their energy-momentum in the tube regions is negligible compared to the integrated energy-momentum in the bulk for large enough $n$.\footnote{Such an argument only makes sense for terms in $V_E$ with support on the full compact space and therefore does not seem to apply to the O6-plane term in \eqref{pot}, which is localized precisely in the region where our solution breaks down. However, according to \eqref{pot2}, the term is equivalent to a bulk term for which our reasoning applies.}
We therefore expect that the scalar potential of the full backreacted solution is well approximated by the smeared one, as emphasized in Section \ref{sec:scalar1}.
The scalings of $V_E$ and its various corrections are summarized in Table \ref{tab-corr}.

\begin{table}[t]\renewcommand{\arraystretch}{1.2}\setlength{\tabcolsep}{10pt}
\centering
\begin{tabular}{|l|l|}
\hline 
correction to $V_E$ & order \\
\hline 
leading potential & $n^{-9/2}$ \\
backreaction in the bulk & $\lesssim n^{-5}$ \\
$\alpha^\prime$ in the bulk (8-derivative) & $\lesssim n^{-5}$ \\
$\alpha^\prime$, $g_s$, backreaction in tube regions & $\lesssim n^{-21/4}$\\
$g_s$ in the bulk (8-derivative, 1-loop) & $\lesssim n^{-13/2}$ \\
\hline
\end{tabular} 
\caption{Corrections to the 4d scalar potential compared to the leading term.}

\label{tab-corr}
\end{table}

Of course,
the above is only an estimate, and it would clearly be important to understand better how string theory resolves O-plane singularities in the presence of Romans mass.
We have argued that the local solution
near the O-planes, whatever it looks like, does not affect the 4d low-energy physics predicted by the smeared solution in the large-$n$ limit.
However, we cannot exclude that string theory somehow avoids our arguments (and, therefore, our conclusions regarding moduli stabilization and scale separation), which are based on a pure supergravity analysis.

As a final remark, we note that the logic discussed in this section does not apply to some dS solutions with O8-planes recently proposed in \cite{Cordova:2018dbb}. As explained above, the 4d low-energy physics of a trustworthy string compactification should not be sensitive to the unknown local physics near the O-planes. The leading contribution to 4d observables such as the cosmological constant should therefore come from the known (supergravity) part of the calculation, while corrections---although perhaps relevant locally near the O-planes---should be subleading. We argued above that this is indeed the case in the DGKT solutions and their non-supersymmetric cousins because $V_E \gg \delta V_E$ at large $n$. Therefore, even though we cannot trust our 10d solution everywhere on the internal space, we can trust the fact that $\langle V_E\rangle < 0$, i.e., that the potential admits AdS solutions. On the other hand, in the setup proposed in \cite{Cordova:2018dbb}, the cosmological constant vanishes when calculated in supergravity \cite{Cribiori:2019clo}.\footnote{This can be shown to follow if one assumes standard boundary conditions for the localized sources in supergravity. See \cite{Cordova:2019cvf} for recent doubts about this assumption.} This implies that the positive cosmological constant found in \cite{Cordova:2018dbb} is entirely generated by effects from the unreliable stringy regions, $\langle V_E\rangle \sim \delta V_E$, such that we cannot trust its sign. Of course, this does not necessarily mean that these solutions do not exist. However, in contrast to the AdS solutions studied in this paper, they cannot be established without a precise knowledge of the local physics near the O-planes.
\\

\section{AdS/KK Scale Separation in String Theory}
\label{sec:scale}

In this section, we will address the question whether string theory admits AdS solutions with a parametric separation between the AdS and KK scales. The existence of the backreacted DGKT solutions and their non-supersymmetric cousins suggests that this is indeed the case. On the other hand, many AdS solutions of string theory are known to not have any scale separation. In Section \ref{sec:scale-review}, we will briefly discuss previous work concerned with this phenomenon, in particular the (strong) AdS distance conjecture of \cite{Lust:2019zwm} and an argument against scale separation in \cite{Gautason:2015tig}. In Section \ref{sec:scale-obs}, we will then study scaling symmetries of the supergravity equations and argue that they explain why there is no scale separation in most AdS compactifications of string theory.\footnote{Aside from the KK tower, other towers of light states may arise in certain string vacua. Whether the AdS scale can be parametrically separated from such other scales will not be discussed here.}
For concreteness, we will focus on type II string theory in the following.

\subsection{Review of Previous Work}
\label{sec:scale-review}

Recently, there has been a renewed interest in the question whether string theory admits AdS solutions with scale separation (see, e.g., \cite{Gautason:2015tig, Lust:2019zwm, Blumenhagen:2019vgj, Font:2019uva}).
In \cite{Lust:2019zwm}, it was conjectured that AdS vacua in the limit $\Lambda\to 0$ satisfy
\begin{equation}
M \sim |\Lambda|^\alpha \label{adc1}
\end{equation}
in Planck units, where $M$ is the mass scale of an infinite tower of states (such as the KK tower), $\Lambda$ is the $d$-dimensional cosmological constant and $\alpha$ is an $\mathcal{O}(1)$ parameter. More specifically, the strong version of the conjecture states that\footnote{It was recently argued that the conjecture receives quantum corrections \cite{Blumenhagen:2019vgj}. Such corrections will not be relevant for the present work, where we only study classical solutions. Furthermore, there is a conjecture that non-supersymmetric AdS solutions are necessarily unstable \cite{Ooguri:2016pdq, Freivogel:2016qwc, Danielsson:2016mtx}. Again, this does not play a role for the following discussion, as our arguments do not rely on stability.}
\begin{align}
& \text{strong AdS distance conjecture:} && \alpha=\frac{1}{2} && (\text{supersymm. AdS$_d$}), \notag \\
& && \alpha=\mathcal{O}(1) && (\text{non-supersymm. AdS$_d$}). \label{ads-conj}
\end{align}
If $M$ is given by the KK scale (as was argued in \cite{Lust:2019zwm} to be the case in string theory), the value $\alpha=\frac{1}{2}$ implies
\begin{equation}
\frac{M_\text{KK}^2}{|\Lambda|} \sim \mathcal{O}(1) \label{no-scale-sep}
\end{equation}
and therefore an absence of scale separation between the KK scale and the AdS curvature scale. Indeed, with $M_\text{KK} \sim 1/R_\text{KK}$ and $\Lambda\sim R_\text{AdS}^{-2}$, we then find $\frac{R_\text{AdS}}{R_\text{KK}} \sim \mathcal{O}(1)$ in the limit $\Lambda\to 0$.
Therefore, assuming the above statements hold in string theory, a necessary condition for scale separation would be broken supersymmetry.

Indeed, it is well-known that many supersymmetric AdS solutions in string theory do not have any scale separation (e.g., \cite{Freund:1980xh, Tsimpis:2012tu, Apruzzi:2019ecr}) and thus support the conjecture \eqref{ads-conj}. However, the DGKT vacua appear to violate \eqref{ads-conj} since they are supersymmetric and exhibit scale separation in the large-$n$ limit. In particular, it was shown in \cite{DeWolfe:2005uu} that they satisfy
\begin{equation}
M_\text{KK} \sim n^{-7/4}, \quad \Lambda \sim n^{-9/2} \label{dgkt-ml}
\end{equation}
such that
\begin{equation}
\alpha = \frac{7}{18}. \label{dgkt-alpha}
\end{equation}
It was pointed out in \cite{Lust:2019zwm} that this is not necessarily a counter-example to \eqref{ads-conj} because the DGKT vacua are smeared and do therefore not solve the 10d equations of motion. However, this criticism does not apply anymore since we showed in this paper that, at large $n$, the smeared DGKT solutions approximate the exact backreacted solutions with arbitrary precision.\footnote{Strictly speaking, we do not know whether terms in the scalar potential that are subleading in $1/n$ break supersymmetry in these solutions. However, even in that case, the corrections would become infinitesimally small at large $n$, i.e., $|DW|^2 \lesssim |W|^2/\sqrt{n}$ (cf.~Section \ref{sec:scalar}). The solutions therefore become supersymmetric in the limit $n\to\infty$.} This suggests that the strong AdS distance conjecture is violated in string theory.

It is also worth pointing out that string theory has \emph{non-supersymmetric} AdS solutions with $\alpha=\frac{1}{2}$ (such as the AdS$_7$ solutions discussed in Appendix \ref{sec:back-ads7}) and with $\alpha\neq \frac{1}{2}$ (such as the non-supersymmetric cousins of the DGKT vacua found in \cite{Marchesano:2019hfb}). Unlike the DGKT vacua, these examples are not in conflict with \eqref{ads-conj}. However, together with the DGKT vacua, they indicate that broken supersymmetry is neither a necessary nor a sufficient condition for scale separation. If this is correct, an interesting question is whether there is a simple criterion other than supersymmetry by which one can distinguish AdS solutions with and without scale separation.

Another work studying the conditions for scale separation in string theory is \cite{Gautason:2015tig} (see also \cite{Petrini:2013ika}).
There, it was argued that classical $d=4$ AdS solutions in type II (and 11d) supergravity cannot have scale separation unless they have either \emph{O-planes} or \emph{large integrated dilaton gradients}.\footnote{More specifically, the requirement on the dilaton gradients was argued to be
$\int \d^6 y \sqrt{g_6}\, w^4(\partial \tau)^2 \gg \sum_q \int \d^6 y \sqrt{g_6}\, w^4 |F_q|^2$ in string frame (plus a further positive contribution on the right-hand side in the presence of D-branes).}
This conclusion was reached in \cite{Gautason:2015tig} by deriving a bound
\begin{equation}
\frac{M_\text{KK}^4}{|\Lambda|} \lesssim \mathcal{O}(1) \label{bound}
\end{equation}
in Planck units for all AdS solutions that do not satisfy either of the two mentioned requirements.\footnote{In the notation of \cite{Gautason:2015tig}, the ratio appearing in the inequality is $M_\text{KK}^4/(M_p^2M_\Lambda^2)$.} However, on closer inspection, this bound does not necessarily forbid AdS/KK scale separation: Indeed, \eqref{bound} does not imply \eqref{no-scale-sep} but only the weaker condition $M_\text{KK}^2/|\Lambda| \lesssim M_\text{KK}^{-2}$. The latter is compatible with AdS/KK scale separation (i.e., with $M_\text{KK}^2/|\Lambda| \gg 1$) whenever the KK scale is small in Planck units (which is true in a controlled flux compactification).\footnote{In the real world, $|\Lambda|$ is tiny such that $M_\text{KK}^4/|\Lambda| \gg 1$, i.e., there is a separation between the vacuum energy scale and the KK scale. However, the relevant condition for the consistency of the 4d low-energy theory in the context of AdS compactifications is $M_\text{KK}^2/|\Lambda| \gg 1$, i.e., a separation between the curvature scale and the KK scale.}
For example, in the DGKT vacua, one finds $M_\text{KK}^4/|\Lambda| \sim n^{-5/2}$ but $M_\text{KK}^2/|\Lambda| \sim n$, where the latter can be made arbitrarily large by increasing $n$.
In the language of the AdS distance conjecture, \eqref{bound} translates to $\alpha \ge \frac{1}{4}$ in the limit $\Lambda\to 0$, while an absence of scale separation would mean $\alpha=\frac{1}{2}$.

A further caveat is that the derivation of the bound \eqref{bound} in \cite{Gautason:2015tig} relies on the estimate $M_\text{KK}^{-4}\sim |\int R_6|$, where $|\int R_6|$ is a weighted integral over the internal Einstein-frame curvature which corresponds to $|\int R_6| \equiv \left|\int \d^6 y \sqrt{g_6}\, w^4\left[\tau^2 R_{mn}g^{mn}+(\partial\tau)^2-\frac{7}{2}\tau\nabla^2\tau\right] \right|$ in our string-frame conventions. As pointed out in \cite{Gautason:2015tig}, this estimate need not be accurate on general manifolds. A simple counter-example would be the smeared DGKT vacua, which have a finite KK scale but $|\int R_6|=0$.\footnote{As a counter-example involving a curved manifold, one may imagine slightly deforming a torus such that it obtains an infinitesimally small curvature while keeping its volume (and, hence, the KK scale) fixed. Taking the dilaton constant for simplicity, one then finds $M_\text{KK}^{-4} \gg |\int R_6| \neq 0$. Note that this is a purely geometric counter-example, i.e., we do not claim that it is a solution of the supergravity equations. In order to decide whether $M_\text{KK}^{-4}\sim |\int R_6|$ holds in the backreacted DGKT vacua, one would have to compute second-order corrections to the metric and the dilaton, as the leading terms in the integrand on the right-hand side can be shown to be a total derivative.}

We conclude that, from what is currently known in the literature, it is not fully clear under which conditions string theory admits vacua with AdS/KK scale separation.
We therefore consider it worthwhile to revisit this issue here.
In particular, we will see in the following section that scale separation is intimately related to scaling symmetries of the classical supergravity equations.
\\

\subsection{An Observation}
\label{sec:scale-obs}

We first consider AdS solutions without O-planes. The key observation is that the type II supergravity equations in this case are invariant \cite{Witten:1985xb, Burgess:1985zz} under the following rescalings of the (string-frame) fields and the D$p$-brane numbers $N_\text{D$p$}$:
\begin{equation}
\tau \sim m h^{1/2}, \quad w \sim h^{1/2}, \quad g_{mn} \sim h, \quad F_q \sim m h^{q/2}, \quad H_3 \sim h, \quad N_\text{D$p$} \sim mh^{(8-p)/2} \label{scalings0}
\end{equation}
for arbitrary parameters $m$, $h$.
Any solution to the supergravity equations therefore comes as a two-parameter family. Note that $m$, $h$ are often related to quantized flux numbers and therefore discrete parameters (unless $F_{q}$ and $H_3$ are exact or zero). We also observe that, for large $h$ and $m h^{1/2}$, the solutions are under perturbative control, i.e., at large volume and small $g_s$.\footnote{Since the large volume is accompanied by large fluxes for $h\to\infty$, we should check that the string-frame curvatures/field strengths actually become small in this limit. This is indeed the case because $R_{mn}g^{mn}\sim |H_3|^2\sim \tau^{-2}|F_q|^2\sim h^{-1}$. Also note that D-branes with $p\neq 3$ typically yield curvature singularities in which case one has to check more carefully where the supergravity solution is trustworthy (cf.~Section \ref{sec:scalar2}).
}

We now compute the dependence of the KK scale and the AdS curvature scale on the two parameters. Using the above scalings, we first find that the $d$-dimensional Planck mass scales like
\begin{equation}
M_p \sim \mathcal{V}_\text{w}^{1/(d-2)} \sim m^{2/(d-2)} h^{5/(d-2)}
\end{equation}
for $d>2$, where $\mathcal{V}_\text{w} = \int \d^{10-d} y \sqrt{g_{10-d}}\, \tau^2 w^{d-2}$ is the warped volume as in Section \ref{sec:scalar}. Computing the exact KK spectrum can be difficult in general as it heavily depends on the details of the geometry.
However, here we are only interested in the parametric scaling. In particular, independent of the details of the compactification, all KK masses should scale inversely with the string-frame volume $\mathcal{V} = \int \d^{10-d} y \sqrt{g_{10-d}}$.
We thus find $M_\text{KK} \sim h^{1/2}/\mathcal{V}^{1/{(10-d)}} \sim m^0h^{0}$ in string units.\footnote{Here, the explicit factor $h^{1/2}$ comes from canonically normalizing the kinetic terms, which are contracted with an inverse spacetime metric $w^{-2}g^{\mu\nu}$ with non-trivial scaling $ \sim h^{-1} $.} The $d$-dimensional scalar curvature scales like $R_d=R_{\mu\nu}g^{\mu\nu} \sim m^0h^{0}$, again in string units.\footnote{Note that there is an ambiguity in the definitions of $M_p$, $M_\text{KK}$ and $R_d$ due to field redefinitions $g_{\mu\nu}\to \gamma^{-2}g_{\mu\nu}$, $w \to \gamma w$ under which the 10d action is invariant. Depending on the chosen convention, one may then obtain different scalings of $M_p$, $M_\text{KK}$ and $R_d$ with respect to $m$ and $h$. In the following, we are only interested in quantities measured in Planck units, for which all rescaling ambiguities cancel out.}
This yields
\begin{align}
\frac{M_\text{KK}}{M_p} & \sim m^{2/(2-d)} h^{5/(2-d)}, \\
\frac{\Lambda}{M_p^d} &\sim \frac{R_d}{M_p^2} \sim m^{4/(2-d)} h^{10/(2-d)}. \label{lambda}
\end{align}
In Planck units, we thus find
\begin{equation}
M_\text{KK} \sim |\Lambda|^{1/2}.
\end{equation}
We have therefore shown that every AdS solution to the supergravity equations with $d>2$ belongs to an infinite two-parameter family which satifies $\alpha=\frac{1}{2}$.\footnote{A priori, the above result would also apply to dS solutions. However, it is well-known that the supergravity equations do not allow such solutions in the absence of O-planes \cite{Gibbons:1984kp, deWit:1986xg, Maldacena:2000mw}.}
Further recall that we require large $h$ and $m h^{1/2}$ for perturbative control. According to \eqref{lambda}, this corresponds to the limit $\Lambda\to 0$ in Planck units.
Our result thus explains why most AdS solutions in string theory satisfy the strong AdS distance conjecture (cf.~\eqref{adc1}, \eqref{ads-conj}). Note, however, that we did not impose supersymmetry in the above derivation. The argument therefore also applies to non-supersymmetric AdS solutions. Furthermore, we did not impose any restriction on the magnitude of the dilaton gradients.

We stress that the above argument is not a proof that \emph{every} path in parameter space will satisfy $\alpha=\frac{1}{2}$ as $\Lambda\to 0$. What we have shown is that there is always at least a two-parameter family of such paths for every AdS solution. However, we cannot exclude that there are solutions with, say, 3 free parameters $m$, $h$, $l$ such that the limits $mh^{1/2},h\to \infty$ and $l\to\infty$ correspond to taking $\Lambda\to 0$ with different values for $\alpha$. Nevertheless, in many known cases, the two parameters $m$, $h$ identified here are in fact the only existing ones. The parameter space spanned by $m$ and $h$ then equals the full parameter space characterizing the corresponding family of AdS vacua, and we have seen that no scale separation occurs there.

We now consider what happens in the presence of O-planes. In order for the two scaling symmetries \eqref{scalings0} to remain unbroken in this case, the O$p$-plane numbers would have to scale like $N_\text{O$p$} \sim N_\text{D$p$} \sim m h^{(8-p)/2}$ (since O-planes contribute to the supergravity equations like D-branes, just with the opposite sign). However, unlike D-branes, O-planes cannot be stacked and therefore appear with a fixed number in a given compactification (i.e., the number of fixed points of the orientifold involution).
Therefore, only rescalings keeping $N_\text{O$p$}$ invariant, i.e., those satisfying $m=h^{(p-8)/2}$, remain a symmetry of the supergravity equations.\footnote{In the presence of O-planes with several different $p$, the scaling symmetries are completely broken.}

However, one checks that, for $p<7$, such rescalings do not lead to AdS solutions in a regime of perturbative control.
Indeed, taking $m=h^{(p-8)/2}$ in \eqref{scalings0}, we find $g_{mn}\sim h$ and $\tau \sim h^{(p-7)/2}$. Hence, we cannot use the scaling parameter $h$ to make the volume parametrically large and the string coupling parametrically small at the same time. Furthermore, flux quantization is a potential issue because $H_3 \sim h$, while $F_0\sim h^{(p-8)/2}$. Finally, recall that the AdS distance conjecture is a statement about vacua in the $\Lambda\to 0$ limit. Using $m=h^{(p-8)/2}$ in \eqref{lambda}, we find that $\Lambda \sim h^{2(p-3)/(2-d)}$ in Planck units.
Depending on $p$, the AdS distance conjecture would therefore either apply for $h\to\infty$ ($p>3$) or for $h\to 0$ ($p<3$), but these are precisely the limits in which the discussed control and quantization issues are most relevant.

To summarize, in the presence of O$p$-planes, at least one combination of the two scaling symmetries $m$ and $h$ is broken, while the orthogonal combination for $p<7$ does not yield parametrically controlled AdS vacua that would allow us to test scale separation and the AdS distance conjecture.\footnote{AdS vacua with O7-planes or O8-planes would avoid this issue, but we are not aware of explicit examples of this type. Some recently proposed non-supersymmetric AdS vacua with O8-planes \cite{Cordova:2018eba} were argued to be incompatible with the supergravity equations in \cite{Cribiori:2019clo}.}

This applies in particular to the DGKT vacua and their non-supersymmetric cousins, which have O$p$-planes with $p=6$. If this was the end of the story, we would have to conclude that these vacua have $\mathcal{O}(1)$ values for the string coupling and the volume and do therefore not exist in a trustworthy regime.
Interestingly though, we have seen in Section \ref{sec:setup-smeared} that a further scaling symmetry arises in these solutions, labelled by the parameter $n$. Crucially, this symmetry differs from the two universal symmetries discussed above. It only arises in the specific setup of Calabi-Yau orientifolds because of the vanishing of certain terms in the supergravity equations. In particular, recall that the smeared solutions are Ricci-flat and have constant/harmonic fields, cf.~\eqref{smeared0}. This implies that the supergravity equations are invariant under the rescalings \eqref{scalings}, which we repeat here for convenience:
\begin{equation}
\tau \sim n^{3/4}, \quad w \sim n^{3/4}, \quad g_{mn} \sim n^{1/2}, \quad F_q \sim n^{q/4}, \quad H_3 \sim n^0. \label{scalings2}
\end{equation}
As explained in Section \ref{sec:scale-review}, these scalings yield $\alpha = \frac{7}{18}$
and therefore a parametric AdS/KK scale separation for $n\to\infty$.\footnote{Such an extra symmetry actually arises in any Ricci-flat compactification with O$p$-planes and constant/harmonic fields.
One verifies that this yields $M_\text{KK}/M_p \sim n^{(p+1)/(2(2-d))}$, $\Lambda/M_p^d \sim n^{(p+d-1)/(2-d)}$ and therefore  $\alpha = \frac{p+1}{2(p+d-1)}$ whenever $\Lambda \neq 0$. Scale separation thus occurs for all such vacua with $d>2$. However, to the best of our knowledge, the smeared DGKT solution, its non-supersymmetric cousins and several smeared Minkowski solutions \cite{Blaback:2010sj} are the only known solutions falling into this class, where the latter have $\Lambda =0$ and therefore no well-defined $\alpha$.}
In the backreacted solution, the symmetry \eqref{scalings2} is broken by subleading effects (cf.~\eqref{ansatz0}--\eqref{ansatz}) but restored in the large-$n$ limit. Asymptotically, we therefore again find $\alpha = \frac{7}{18}$.

We have thus found a simple explanation for the fact that AdS/KK scale separation occurs in the DGKT vacua and their non-supersymmetric cousins,
while it does not in most other AdS solutions in string theory. Indeed, this is related to the different scaling symmetries arising in these solutions, from which one can immediately deduce the asymptotic behavior of $\Lambda$ and $M_\text{KK}$ in the corresponding parameter spaces.
It would be very interesting to understand more generally under which conditions flux vacua can have such extra scaling symmetries beyond the universal ones $m$ and $h$. Since all known vacua with scale separation involve O-planes, one may speculate that the latter might be a necessary condition, as already proposed in \cite{Gautason:2015tig}. We leave a more systematic study of these questions for future work.
\\

\section{Conclusions}

\label{sec:concl}

In this paper, we computed the O6-plane backreaction in type IIA AdS flux vacua on Calabi-Yau orientifolds, which were previously studied in the smeared approximation in several other works \cite{DeWolfe:2005uu, Derendinger:2004jn, Camara:2005dc, Acharya:2006ne,  Narayan:2010em, Marchesano:2019hfb}. We performed our analysis in the usual regime of large volumes and small $g_s$, where string theory is well-approximated by 10d supergravity. Like the smeared vacua, the backreacted solutions we constructed plausibly allow full moduli stabilization, parametric control over string corrections and a parametric separation between the AdS and KK scales. Our results apply equally to supersymmetric and non-supersymmetric AdS solutions that exist in the same setup of type IIA Calabi-Yau orientifolds. As an explicit example illustrating our general results, we studied in detail the backreaction for an orientifold of $T^6/\mathbb{Z}_3^2$.

Along the way, we also clarified the validity of the smeared approximation in these vacua. A recurring criticism in the literature is that smeared O-planes do not make sense because O-planes are intrinsically localized objects (i.e., the fixed points of the orientifold involution) which cannot exist as a smeared configuration. Indeed, a smeared source differs from a localized one in the sense that one can always distinguish the two by performing a sufficiently precise scattering experiment.
However, this objection
misses the main point of smearing. Although a smeared O-plane is not a physical object, the smeared solution may still be useful as an approximation to the low-energy physics of the orientifold compactification.
In particular, a 4d observer cannot distinguish a source localized at some submanifold in the internal space from a smeared source because the relevant length scales cannot be resolved at energies below the KK scale.
The only measurable effect are then the integrated backreaction corrections to the 4d observables.
The common lore is therefore that the smeared solution should approximate the 4d low-energy EFT
at large volumes and small $g_s$ since the near-source regions with relevant backreaction are then very small compared to the overall volume. Our results made these heuristic ideas
precise in an explicit computation.
Indeed, we showed that, at large $n$, the backreaction corrections to the smeared solution are parametrically suppressed almost everywhere on the internal space. This in particular implies that the 4d scalar potential equals the smeared expression in the large-$n$ limit, up to subleading $1/\sqrt{n}$ corrections.

For future work, it would be interesting to translate our results into the language of 4d $\mathcal{N}=1$ supergravity, i.e., to compute the NLO backreaction corrections to the K\"{a}hler potential derived in \cite{Grimm:2004ua}. It would also be interesting to study in more detail the geometry of the backreacted internal spaces, which receive curvature corrections and are therefore not Calabi-Yau. For the supersymmetric case, this was argued in \cite{Saracco:2012wc} to yield $\text{SU}(3)\times \text{SU}(3)$ structure manifolds.

Furthermore,
an important question is
how string theory resolves O-plane singularities in the presence of Romans mass where no M-theory lift is available. An interesting proposal for such a resolution mechanism was made in \cite{Saracco:2012wc}. It was argued in \cite{Gautason:2015tig} that such a mechanism cannot be realized in the DGKT vacua. However, this argument was based on the assumption that vacua with scale separation have to violate the bound \eqref{bound},
which we showed not to be the case for sufficiently small KK scale. It therefore remains an open question whether the O6-plane singularities in the DGKT vacua are resolved in a similar way as in the local solutions of \cite{Saracco:2012wc}.
Understanding the local physics
in the near-O-plane regions would in any case be important to see whether there are obstructions to some of the conclusions drawn in this paper.
For example, it would be important to see whether light states somehow arise in the stringy solution that are not visible in a supergravity analysis.

Aside from the DGKT vacua and their non-supersymmetric cousins, our results are also relevant for the relation between smeared and backreacted solutions in general. To our knowledge, only very few flux compactifications are known in which the full D-brane/O-plane backreaction is taken into account (e.g., \cite{Giddings:2001yu, Dasgupta:1999ss, Apruzzi:2013yva}). Our method allows to compute the backreaction in general, provided the setup has a smeared solution which admits a large-volume/small-$g_s$ limit. In that case, the leading-order solution is expected to equal the smeared one, and subleading corrections are given as solutions to linear equations with Dirac-delta sources.
Note that, as discussed in more detail in Appendix \ref{sec:back}, this reasoning only works for vacua where the large-volume/small-$g_s$ limit can be taken while keeping the number of localized sources \emph{fixed}. Importantly, this subtlety implies that, for some flux vacua, smearing is never a good approximation at arbitrarily large volumes and small $g_s$. The usual slogan that backreaction becomes negligible in this limit should therefore be taken with a grain of salt.

Nevertheless, we expect that our method can be useful in many cases, and it would be very interesting to apply it to further Minkowski or AdS solutions that are only known as smeared solutions so far.
For example, there are type IIB AdS vacua with smeared O5/O7-planes which were argued to admit a large-volume/small-$g_s$ limit and parametric AdS/KK scale separation \cite{Caviezel:2008ik, Caviezel:2009tu, Petrini:2013ika}. Using the techniques developed in this paper, it should be possible to compute the backreaction corrections to these smeared solutions.
One might furthermore be tempted to use similar techniques to compute the backreaction in classical dS solutions. However, unlike Minkowski or AdS vacua, classical dS solutions do probably not exist in a regime of parametrically large volumes and small $g_s$ \cite{Junghans:2018gdb, Banlaki:2018ayh}. This indicates that they cannot be studied in terms of a systematic expansion around the smeared solution.

Finally, we studied the strong AdS distance conjecture of \cite{Lust:2019zwm}, which posits that a parametric scale separation is impossible for supersymmetric AdS vacua in the limit $\Lambda \to 0$. Our results suggest that the conjecture is violated in string theory by the backreacted DGKT solutions.
We also revisited an earlier result in \cite{Gautason:2015tig} where a no-go argument against scale separation was proposed for AdS vacua without O-planes. We found that the bound derived in \cite{Gautason:2015tig} does not exclude a parametric AdS/KK scale separation if the KK scale is sufficiently small compared to the Planck scale. The result of \cite{Gautason:2015tig} further assumes a specific relation between the KK scale and the internal curvature which need not hold on general manifolds. We therefore presented a different argument in this paper, which does not rely on these two assumptions. In particular, we observed that scaling symmetries of classical supergravity explain why most AdS solutions in string theory do not exhibit scale separation and why the DGKT vacua and their non-supersymmetric cousins behave differently.
It would be very interesting to extend this simple argument to a complete classification of flux vacua with and without scale separation in string theory.
\\

\section*{Acknowledgments}

I would like to thank David Andriot and Thomas Van Riet for helpful correspondence.
\\

\appendix

\section{Backreaction in General}
\label{sec:back}

As stated before, our method to compute the backreaction should be applicable not only in the DGKT setup but quite generally in many families of flux vacua that admit a large-volume/small-$g_s$ limit. In this limit, the backreaction of a brane or O-plane is expected to become small such that the exact solution is the smeared one plus a subleading correction, which is determined by a set of linear equations as in Section \ref{sec:sol}. We will demonstrate this in Appendix \ref{sec:back-gkp} using the familiar example of the GKP vacua of type IIB supergravity. Of course, the backreacted solution is already known in this case, so that our result in this section is not new. The reason we discuss the GKP example is because it is simple and nicely illustrates how our method generalizes to flux vacua other than the DGKT solutions. Indeed, we find that the result obtained with our method agrees precisely with the known expressions as expected.

A second point we want to emphasize is that taking the large-volume/small-$g_s$ limit can be subtle sometimes and does \emph{not always} imply that the backreaction becomes negligible. This point has led to some confusion in the literature since it means that there are types of flux vacua where smearing is justified and types of flux vacua where it is not. Indeed, in Appendix \ref{sec:back-ads7}, we will discuss a family of AdS$_7$ flux vacua in type IIA which admit a large-volume/small-$g_s$ limit but the limit cannot be taken in such a way that the backreaction becomes small. The reason is that the decreasing backreaction of the individual branes is compensated by an increasing brane number as we take the limit. The total backreaction therefore remains a leading-order effect for all volumes. Consequently, these solutions are a counter-example where smearing does not become parametrically good and our method cannot be applied.

\subsection{GKP Vacua}
\label{sec:back-gkp}

We first discuss the GKP solutions. These are Minkowski flux vacua of type IIB supergravity on (conformally) Calabi-Yau orientifolds \cite{Giddings:2001yu,Dasgupta:1999ss}. For simplicity, we focus on solutions with O3-planes as the only localized sources.
The string-frame metric is given by
\begin{equation}
\d s_{10}^2 = w^2 \eta_{\mu\nu}\d x^\mu \d x^\nu + g_{mn}\d y^m \d y^n, \qquad g_{mn}=w^{-2} \tilde g_{mn}, \label{gkp-metric}
\end{equation}
where $\eta_{\mu\nu}$ is the Minkowski metric, $\tilde g_{mn}$ is Ricci-flat and $w=w(y)$ is the warp factor as usual. The NSNS and RR field strengths satisfy (again in string frame)
\begin{equation}
F_1=0, \quad F_3 = - \tau \star_6 H_3, \quad F_5 = - (1+\star_{10}) \tau \star_6 \d w^{4}, \label{gkp-form}
\end{equation}
where $\d H_3=\d F_3=0$.\footnote{The 10d and 6d Hodge stars are defined with respect to the full metric including the warp factors.} One can show that, with this ansatz, all equations of motion are solved provided the dilaton is a constant (which is fixed in terms of the 3-form fluxes) and the warp factor satisfies \cite{Giddings:2001yu,Giddings:2005ff}
\begin{equation}
w(y)^{-4} = \beta(y) + c, \label{wgkp}
\end{equation}
where $c$ is an arbitrary constant and the function $\beta(y)$ is a solution of the Poisson equation\footnote{The tildes indicate that the implicit inverse metrics and metric determinants are built from $\tilde g_{mn}$, i.e., there is no implicit dependence on warp factors in this equation.}
\begin{equation}
\tilde\nabla^2 \beta = - |\tilde H_3|^2 + \frac{1}{4\tau}\sum_i \frac{\delta^{(6)}(y-y_i)}{\sqrt{\tilde g_6}}. \label{aeq}
\end{equation}
The solution thus has a free parameter $c$ which shifts the warp factor by a constant. Substituting \eqref{wgkp} into \eqref{gkp-metric}, we observe that $c$ is (up to normalization) simply the volume modulus. We therefore expect that the backreaction of the O3-planes becomes negligible in the large-$c$ limit almost everywhere on the internal space.

In the smeared solution, the delta distributions in \eqref{aeq} are replaced by a constant which cancels the flux term such that $\tilde \nabla^2 \beta = 0$ \cite{Blaback:2010sj}. The non-constant part $\beta$ in the warp factor is therefore absent, and we can write
\begin{equation}
w=c^{-1/4},
\end{equation}
where all other fields are given in terms of $w$ as above. In the smeared solution, the volume modulus is therefore a rescaling mode with
\begin{equation}
\tau \sim c^{0}, \quad w \sim c^{-1/4}, \quad g_{mn} \sim c^{1/2}, \quad F_3, H_3 \sim c^0.
\end{equation}

It is now straightforward to see that the localized solution \eqref{wgkp} approaches the smeared one at large $c$ with
\begin{equation}
w = c^{-1/4} + c^{-5/4} w^{(1)} + \mathcal{O}(c^{-9/4}) \label{wgkp2}
\end{equation}
(where $\beta=-4w^{(1)}$), and analogously for the other fields. Had we not known the localized solution already, we could therefore have computed it in terms of a $1/c$ expansion of the fields around the smeared solution as in Section \ref{sec:sol}. Note that it superficially looks as if the result \eqref{wgkp} is more general than \eqref{wgkp2}, as it seems to hold for any $c$. However, this is not the case. Indeed, string theory is not approximated by supergravity at small volumes. Computing the backreaction using the supergravity equations is therefore only meaningful at large $c$.
Close to the O3-planes, the backreaction becomes large and the $1/c$ expansion breaks down. However, the non-linear solution would not be reliable in this region either, as string effects become locally relevant there.
\\

\subsection{AdS$_7$ Flux Vacua with D6-branes}
\label{sec:back-ads7}

It is instructive to also discuss a counter-example where our method does not work. We consider AdS$_7$ flux vacua of massive type IIA string theory on a conformal $S^3$. The solutions have $H_3$ flux on the 3-sphere and exist for various configurations of spacetime-filling D6/D8-branes. For simplicity, we will focus on a family of vacua with a single stack of $N_\text{D6}$ D6-branes. The corresponding smeared solution and some local and global properties of localized solutions were found in \cite{Blaback:2011nz, Blaback:2011pn}. The localized solution was first constructed numerically in \cite{Apruzzi:2013yva}, and the analytic solution was discovered in \cite{Apruzzi:2015zna}.
A remarkable fact is that the smeared solution is non-supersymmetric \cite{Blaback:2011nz, Danielsson:2013qfa}, while there are both supersymmetric \cite{Apruzzi:2013yva} and non-supersymmetric \cite{Dibitetto:2015bia, Passias:2015gya, Apruzzi:2016rny} localized solutions. This already indicates that the smeared solution is not a good approximation to the localized ones.

One can show that the solutions come in families with two free parameters $h$ and $m$, which are associated to the $H_3$ flux on the 3-sphere and the Romans mass parameter. The various fields scale like \cite{Junghans:2014wda}
\begin{align}
&\tau \sim m h^{1/2}, \quad w \sim h^{1/2}, \quad g_{mn} \sim h, \quad F_2 \sim mh, \quad F_0 \sim m, \quad H_3 \sim h, \quad N_\text{D6} \sim mh.
\end{align}
Hence, in the large-$h$ limit, we are at large volumes and small $g_s$. However, crucially, taking the large-$h$ limit also involves a rescaling of the brane number $N_\text{D6}$. The common lore that the backreaction becomes negligible in this limit does therefore not apply here. Although the backreaction of an individual D6-brane becomes indeed small at large $h$, the brane number increases in such a way that the total backreaction remains a leading-order effect for any value of $h$. Unlike in the DGKT or GKP vacua, there is therefore no limit in which the exact solution is approximated by the smeared one plus a subleading correction.

For example, using the results of \cite{Blaback:2011nz, Apruzzi:2015zna, Passias:2015gya}, one can compute the 7d cosmological constants of the smeared solution and the localized solutions (for the same flux quanta). This yields (in Planck units)\footnote{Here, we normalized $m$ and $h$ such that $mh=N_\text{D6}$ and $m=F_0$.}
\begin{align}
\Lambda_\text{smeared} &= -\frac{125}{49} \frac{7^{3/5}\pi^{14/5}}{m^{4/5}h^2} \approx - \frac{202.2}{m^{4/5}h^2}, \\ \Lambda_\text{loc,SUSY} &= -\frac{15}{2} \frac{2^{1/5}3^{4/5}5^{2/5}\pi^{6/5}}{m^{4/5}h^2} \approx - \frac{156.0}{m^{4/5}h^2}, \\
\Lambda_\text{loc,non-SUSY} &= \frac{2^{8/5}}{3} \Lambda_\text{loc,SUSY} \approx - \frac{157.6}{m^{4/5}h^2}.
\end{align}
Note that the three results differ by the same relative factor for all $m$, $h$ such that the smeared approximation is never parametrically good. This was different in Section \ref{sec:scalar}, where we found that the scalar potential in DGKT (and, hence, the cosmological constant) equals the smeared one at large $n$ up to subleading terms.
By contrast, the smeared solution in the present case should not be viewed as an approximation to the localized solutions but rather as a physically different solution in which the branes are not localized on one of the poles but spread over the whole sphere.
\\

\bibliographystyle{utphys}
\bibliography{groups}

\end{document}